\documentclass[11pt,a4paper]{article}

\usepackage[numbers, square, comma, sort&compress]{natbib}
\usepackage{amssymb}
\usepackage{amsthm}
\usepackage{amstext}
\usepackage{amsmath}
\usepackage{amsfonts}
\usepackage{empheq}
\usepackage{verbatim}
\usepackage{geometry}
\usepackage{latexsym}
\usepackage{t1enc}
\usepackage{graphicx}
\usepackage[all]{xy}
\usepackage{makeidx}
\usepackage{slashed}
\usepackage{multicol}
\usepackage{alltt}

\allowdisplaybreaks

\newcommand{\be}{\begin{equation}}
\newcommand{\ee}{\end{equation}}
\newcommand{\bea}{\begin{eqnarray}}
\newcommand{\eea}{\end{eqnarray}}

\def\beq{\begin{equation}}
\def\eeq{\end{equation}}
\def\beqa{\begin{eqnarray}}
\def\eeqa{\end{eqnarray}}

\def\spa#1.#2{\left\langle#1\,#2\right\rangle}
\def\spb#1.#2{\left[#1\,#2\right]}
\def\spash#1.#2{\spa{\smash{#1}}.{\smash{#2}}}
\def\spbsh#1.#2{\spb{\smash{#1}}.{\smash{#2}}}
\def\sand#1.#2.#3{%
  \left\langle\smash{#1^{-}}{\vphantom1}\right|{#2}%
  \left|\smash{#3^{-}}{\vphantom1}\right\rangle}
\def\sandp#1.#2.#3{%
  \left\langle\smash{#1^{-}}{\vphantom1}\right|{#2}%
  \left|\smash{#3^{+}}{\vphantom1}\right\rangle}
\def\sandpp#1.#2.#3{%
  \left\langle\smash{#1^{+}}{\vphantom1}\right|{#2}%
  \left|\smash{#3^{+}}{\vphantom1}\right\rangle}
\def\sandpm#1.#2.#3{%
  \left\langle\smash{#1^{+}}{\vphantom1}\right|{#2}%
  \left|\smash{#3^{-}}{\vphantom1}\right\rangle}
\def\sandmp#1.#2.#3{%
  \left\langle\smash{#1^{-}}{\vphantom1}\right|{#2}%
  \left|\smash{#3^{+}}{\vphantom1}\right\rangle}

\def\ssand#1.#2.#3{%
  \left\langle\smash{#1}{\vphantom1}\right|{#2}%
  \left|\smash{#3}{\vphantom1}\right]}
\def\ssandp#1.#2.#3{%
  \left\langle\smash{#1}{\vphantom1}\right|{#2}%
  \left|\smash{#3}{\vphantom1}\right\rangle}
\def\ssandpp#1.#2.#3{%
  \left\langle\smash{#1}{\vphantom1}\right|{#2}%
  \left|\smash{#3}{\vphantom1}\right\rangle}

\def\proj{\flat}
\def\projdot#1.#2{k_{#1}^\proj\cdot k_{#2}^\proj}
\def\sandff#1.#2.#3{%
  \left\langle\smash{#1^{\proj,-}}{\vphantom1}\right|{#2}%
  \left|\smash{#3^{\proj,-}}{\vphantom1}\right\rangle}
\def\sandnf#1.#2.#3{%
  \left\langle\smash{#1^{-}}{\vphantom1}\right|{#2}%
  \left|\smash{#3^{\proj,-}}{\vphantom1}\right\rangle}
\def\sandfn#1.#2.#3{%
  \left\langle\smash{#1^{\proj,-}}{\vphantom1}\right|{#2}%
  \left|\smash{#3^{-}}{\vphantom1}\right\rangle}

\def\spa#1.#2{\left\langle#1\,#2\right\rangle}
\def\spb#1.#2{\left[#1\,#2\right]}

\numberwithin{equation}{section}

\begin{document}


\begin{titlepage}

\hbox{Edinburgh 2017/13}
\hbox{CERN-TH-2017-138}
\hbox{CP3-17-20}
\hbox{SLAC-PUB-17089}
\hbox{QMUL-PH-17-10}

\vskip 25mm

\begin{center}
\Large{\sc{Bootstrapping the QCD soft anomalous dimension}}
\end{center}

\vskip 8mm

\begin{center}
\O{}yvind Almelid$^{a}$~\footnote{almelid@pvv.ntnu.no},
Claude~Duhr$^{b,c}$~\footnote{claude.duhr@cern.ch},
Einan~Gardi$^a$~\footnote{einan.gardi@ed.ac.uk},
Andrew~McLeod$^{d,e}$~\footnote{ajmcleod@stanford.edu},
Chris~D.~White$^f$~\footnote{christopher.white@qmul.ac.uk} \\ [6mm]

\vspace{6mm}
\textit{$^a$ Higgs Centre for Theoretical Physics, School of Physics and Astronomy, \\
The University of Edinburgh, Edinburgh EH9 3FD, Scotland, UK}\\
\vspace{1mm}

\textit{$^b$ Theoretical Physics Department, CERN, Geneva, Switzerland} \\ 
\vspace{1mm}

\textit{$^c$ Center for Cosmology, Particle Physics and Phenomenology
  (CP3), \\Universit\'{e} Catholique de Louvain, 1348 Louvain-La-Neuve,
  Belgium} \\
\vspace{1mm}

\textit{$^d$ SLAC National Accelerator Laboratory, Stanford University, Stanford, CA 94309, USA}\\
\vspace{1mm}

\textit{$^e$ Kavli Institute for Theoretical Physics, UC Santa Barbara, Santa Barbara, CA 93106, USA}\\
\vspace{1mm}

\textit{$^f$Centre for Research in String Theory, School of Physics and Astronomy, 
Queen Mary University of London, 327 Mile End Road, London E1 4NS, UK} \\
\vspace{1mm}

\end{center}

\vspace{5mm}

\begin{abstract}
\noindent The soft anomalous dimension governs the infrared
singularities of scattering amplitudes to all orders in perturbative quantum
field theory, and is a crucial ingredient in both formal and
phenomenological applications of non-abelian gauge theories. It has
recently been computed at three-loop order for massless partons by explicit evaluation of
all relevant Feynman diagrams. In this paper, we show how the same
result can be obtained, up to an overall numerical factor, using a
bootstrap procedure. We first give a geometrical argument for
the fact that the result can be expressed in terms of single-valued harmonic
polylogarithms. We then use symmetry considerations as well as known properties of scattering amplitudes in collinear and high-energy (Regge) limits to constrain an ansatz of basis functions. This is a highly non-trivial cross-check of the result, and
our methods pave the way for greatly simplified higher-order
calculations.
\end{abstract}

\end{titlepage}


\section{Introduction}
\label{sec:introduction}

The calculation of higher-order perturbative corrections in
non-abelian gauge theories is crucial both for increasing the
precision of collider physics predictions, as well as for our
understanding of field theory itself. To this end, it is important to
understand those quantities which dictate all-order properties of
perturbative scattering amplitudes. One such quantity is the {\it soft
  anomalous dimension}, which governs the long-distance singularities
of scattering amplitudes, and can also be deduced from the ultraviolet
singularities of correlators of Wilson line
operators~\cite{Polyakov:1980ca,Arefeva:1980zd,Dotsenko:1979wb,Brandt:1981kf,Korchemsky:1985xj,Korchemsky:1985xu,Korchemsky:1987wg}. These
divergences have been studied in QCD for many decades for processes
involving two partons and any number of colour singlet
particles~\cite{Mueller:1979ih,Collins:1980ih,Sen:1981sd,Sen:1982bt,Gatheral:1983cz,Frenkel:1984pz,Sterman:1981jc,Magnea:1990zb}. The
singularity structure of amplitudes involving several partons has been
examined more recently, in both the
massless~\cite{Korchemsky:1993hr,Korchemskaya:1996je,Korchemskaya:1994qp,Catani:1996vz,Catani:1998bh,Sterman:2002qn,Dixon:2008gr,Kidonakis:1998nf,Bonciani:2003nt,Dokshitzer:2005ig,Aybat:2006mz,Gardi:2009qi,Becher:2009cu,Becher:2009qa,Gardi:2009zv,
  Dixon:2009gx,Dixon:2009ur,DelDuca:2011xm,DelDuca:2011ae,Caron-Huot:2013fea,Ahrens:2012qz,Naculich:2013xa,Erdogan:2014gha,Gehrmann:2010ue}
and
massive~\cite{Kidonakis:2009ev,Mitov:2009sv,Becher:2009kw,Beneke:2009rj,Czakon:2009zw,Ferroglia:2009ii,Chiu:2009mg,Mitov:2010xw,Gardi:2013saa,Falcioni:2014pka,Henn:2013fah,Dukes:2013gea,Dukes:2013wa,Gardi:2010rn,Gardi:2011wa,Gardi:2011yz,Gardi:2013ita,Laenen:2008gt}
cases. Very recently, a first calculation of the multileg soft
anomalous dimension for massless particles has been presented at
three-loop order~\cite{Almelid:2015jia}. Despite the highly
complicated nature of the relevant Feynman integrals, the final result
is very simple when written in the right way. It consists of a
contribution that is consistent with the so-called {\it dipole
  formula} of refs.~\cite{Gardi:2009qi,Becher:2009cu,Becher:2009qa}
and depends termwise only on pairs of particles, a constant term
depending on the colour charges of sets of three partons, and a
contribution involving sets of four partons that depends on conformally invariant cross ratios of the
invariants $\beta_i\cdot\beta_j$, where $\beta_i$ is the four-velocity
of the $i^{\rm th}$ Wilson line. It was found in
ref.~\cite{Almelid:2015jia} that this last contribution, which can
first appear at three loops, can be written in terms of a restricted
class of functions: single-valued harmonic
polylogarithms~\cite{Brown:2004,Remiddi:1999ew,Dixon:2012yy} (SVHPLs),
whose arguments depend on conformally invariant cross ratios.

The unexpected simplicity of the three-loop massless soft anomalous
dimension calls for a deeper understanding and suggests that one may
obtain it by alternative means, without surrendering to the complexity
of Feynman integral evaluation. In particular, if one knows that the
answer can be expressed exclusively in terms of a restricted set of
functions, then one can use a {\it bootstrap} approach. That is, one
may write an ansatz for the soft anomalous dimension in terms of a
basis of such functions, and constrain the coefficients of this ansatz
by applying known consistency constraints, such as the known behaviour
of the result in certain kinematic limits, e.g. the high-energy limit,
or the limit in which a subset of the Wilson lines becomes
collinear. Such methods have been highly successful in constraining
amplitudes in planar ${\cal N}=4$ Supersymmetric Yang-Mills (SYM)
theory~\cite{Dixon:2012yy,Dixon:2011pw,Dixon:2011nj,Dixon:2013eka,Dixon:2014voa,Dixon:2014iba,Dixon:2015iva,Caron-Huot:2016owq,Drummond:2014ffa,Dixon:2016nkn}.

In the present study, the overall ethos of the bootstrap is very
similar. We start with a detailed argument for why SVHPLs are the only relevant functions. This
need apply only to that part of the soft anomalous dimension depending
on conformally invariant cross ratios, which at three loops is the
most difficult part to compute using Feynman diagrams. Secondly, one
must identify a number of constraints on an ansatz of such
functions. As will be demonstrated, there is a sufficient number of
known constraints at three loops to completely determine the soft
anomalous dimension up to an overall multiplicative rational
factor. As well as known symmetry properties, we will make use of
previously obtained results coming from the
Regge~\cite{DelDuca:2011xm,DelDuca:2011ae,Caron-Huot:2013fea,DelDuca:2013ara,DelDuca:2014cya,Caron-Huot:2017fxr}
and collinear~\cite{Becher:2009qa,Dixon:2009ur} limits. Our results provide new
insights into the structure of the three-loop soft anomalous
dimension, and will also prove useful in investigating this quantity
at higher loop orders.

The structure of the paper is as follows. In section~\ref{sec:review},
we briefly review the definition and known properties of the soft
anomalous dimension. In section~\ref{sec:HPL}, we provide a detailed
argument for why SVHPLs are expected to describe the part of the soft
anomalous dimension depending on conformally invariant cross ratios.
In section~\ref{sec:constraints}, we develop a general ansatz for this
function satisfying the relevant symmetries, and in sections~\ref{sec:Reggelimit}
and~\ref{sec:collinearlimit} we derive constraints on this ansatz based, respectively, on the Regge limit of 2-to-2 scattering and on two-particle collinear limits. In section~\ref{sec:discuss}, we combine these constraints and reproduce completely the form of the non-dipole-like part of the soft anomalous dimension, up to an overall rational number factor. Finally, we discuss our results and conclude. 
Technical details concerning the Regge limit and colour algebra identities are collected in two appendices.

\section{The soft anomalous dimension}
\label{sec:review}

In this section, we review the salient details regarding the soft
anomalous dimension that will be needed for what follows. We start by
considering a scattering amplitude for $n$ massless partons (quarks
or gluons) at {\it fixed angle}, such that all invariants $p_i\cdot
p_j$ are large relative to the QCD confinement scale. It
is well-known that such amplitudes are beset by infrared divergences,
due to the emission of virtual radiation that may become soft and/or
collinear with the external partons. The amplitude, in $d=4-2\epsilon$
spacetime dimensions, then assumes the factorised
form~\cite{Sen:1982bt,Kidonakis:1998nf,Sterman:2002qn,Dixon:2008gr,Aybat:2006mz,Gardi:2009qi}
\begin{equation}
{\cal A}_n(\{p_i\},\epsilon,\alpha_s)=
{\cal S}(\{\beta_i\},\{{\bf T}_i \},\epsilon,\alpha_s)\,
{\cal H}_n(\{p_i\},\{n_i\},\epsilon,
\alpha_s)\,
\prod_{i=1}^n \frac{J(p_i,n_i,\epsilon,\alpha_s)}
{{\cal J}(\beta_i,n_i,\epsilon,\alpha_s)},
\label{softcolfac}
\end{equation}
where $\alpha_s$ is the $d$-dimensional running coupling, ${\cal H}_n$
is a process-dependent {\it hard function} that is finite as
$\epsilon\rightarrow 0$, and ${\cal S}$ and $J$ are the {\it soft} and {\it
  jet functions} that collect infrared singularities originating from
emissions that are soft and collinear to particle $i$
respectively. We have introduced a colour operator ${\bf T}_i$
associated with line $i$, as in ref.~\cite{Catani:1996vz}, which
convert into generators in the appropriate representation when acting
on the hard function. The soft function is thus a complicated object
in colour space, in contrast to the jet functions, which are
colour-singlet objects. The auxiliary vectors $\{n_i\}$, one for each
external parton, are used to define the jet functions in a
gauge-invariant manner, such that their dependence cancels between the jet functions and the hard function. We have suppressed renormalisation
and factorisation scales in eq.~({\ref{softcolfac}) for brevity. The
  functions ${\cal J}$ are referred to as {\it eikonal jet functions},
  and correct for the double counting of contributions which are both
  soft and collinear. The soft and eikonal jet functions depend only
  on the four-velocities $\{\beta_i\}$ (where $\beta_i\propto p_i$) of
  each external parton, and have formal definitions in terms of vacuum
  expectation values of Wilson line operators. The soft function, for
  example, is given by
\begin{equation}
{\cal S}(\{\beta_i\},\{{\bf T}_i\} , \epsilon,\alpha_s)=\left\langle 0\left| 
{\rm T}\left[\Phi_{\beta_1}\,\Phi_{\beta_2}\ldots\Phi_{\beta_n}\right]
\right|0\right\rangle,
\label{Sdef}
\end{equation}
where ${\rm T}[\cdots]$ represents a time-ordered product and 
\begin{equation}
\Phi_{\beta_i}={\cal P}\exp\left[i\mu^\epsilon g_s {\bf
    T}_i^a\int_0^\infty ds\,\beta_i\cdot A^a(s\beta_i)\right]
\label{Phidef}
\end{equation}
is a Wilson-line operator along a trajectory of the $i^{\rm th}$
parton. The soft function is matrix-valued in the space of possible
colour flows in the hard interaction ${\cal H}_n$. The jet functions
also have operator definitions~\cite{Dixon:2008gr}, that will not be
needed in what follows.

The operator defined in eq.~(\ref{Phidef}) is invariant under
reparametrisations of the contour, which translates into an invariance
under rescalings of the four-velocity, $\beta_i\to \kappa_i\beta_i$. This
property, together with Lorentz invariance, dictates that the soft
function for non-lightlike Wilson lines can only depend on the
quantities (related to the {\it cusp angles} between pairs of
Wilson lines)
\begin{equation}
\gamma_{ij}=\frac{2\beta_i\cdot\beta_j+i0}
{\sqrt{\beta_i^2-i0}\,\sqrt{\beta_j^2-i0}},
\label{cuspangle}
\end{equation}
where $i0$ is the usual Feynman prescription, which must be
taken into account when analytically continuing the velocities to
different kinematic regions. Upon regulating all singularities in
eq.~(\ref{softcolfac}), they may be combined into a single overall
factor, such that the amplitude can be written in the alternative
factorised form~\cite{Becher:2009qa,Gardi:2009zv,Gardi:2016ttq}
\begin{equation}
{\cal
  A}_n\left(\{p_i\},\epsilon,\alpha_s(\mu^2)
\right)=
Z_n\left(\{p_i\},\{{\bf T}_i\} ,\epsilon,\alpha_s(\mu_f^2)
\right)
{\cal H}_n\left(\{p_i\},\frac{\mu_f}{\mu}
,\epsilon,\alpha_s(\mu^2)\right).
\label{Andef}
\end{equation}
Here, as in ref.~\cite{Dixon:2009ur}, we have been careful in
distinguishing the ultraviolet renormalisation scale $\mu$ from the
factorisation scale $\mu_f$ at which infrared singularities are
regularised. The complete divergent prefactor $Z_n$ satisfies the
renormalisation group equation
\begin{equation}
\!\frac{d}{d\ln \mu_f}Z_n\left(\{p_i\},\{{\bf T}_i\} ,\epsilon,\alpha_s(\mu_f)\right)=-Z_n\left(\{p_i\},\{{\bf T}_i\} ,\epsilon,\alpha_s(\mu_f)\right)\Gamma_n\left(\{p_i\},\{{\bf T}_i\} ,
\mu_f,\alpha_s(\mu_f)\right),
\label{rengroup}
\end{equation}
which defines the {\it soft anomalous dimension} $\Gamma_n$.  This
will also be matrix-valued in colour flow space, and the solution of
eq.~(\ref{rengroup}) may be formally written as
\begin{equation}
Z_n\left(\{p_i\},\{{\bf T}_i\} ,\epsilon,\alpha_s(\mu_f^2)
\right)={\cal
  P}\exp\left\{-\frac12\int_0^{\mu_f^2}\frac{d\lambda^2}{\lambda^2}
\Gamma_n\left(\{p_i\},\{{\bf T}_i \},\lambda,
\alpha_s(\lambda^2)\right)\right\},
\label{ZnRG}
\end{equation}
where the ${\cal P}$ symbol denotes path ordering of these matrices
according to the ordering of the scales $\lambda$. The soft anomalous
dimension is finite as $\epsilon\rightarrow 0$, such that all
singularities in $Z_n$ are generated by performing the integral over
$\lambda$, and using the known scale dependence of the $d$-dimensional
running coupling. 

Much is known about the structure of $\Gamma_n$. In particular,
explicit calculation of the result for massless partons up to two-loop
order~\cite{Catani:1998bh,Sterman:2002qn,Aybat:2006mz} demonstrated
that potential three-particle correlations were absent at this
order. This motivated a detailed study of constraints on the soft
anomalous dimension, based on the factorisation formula of
eq.~(\ref{softcolfac}), together with invariance under rescaling of
the four-velocities $\{\beta_i\}$. Invariance of the soft function alone
is broken for lightlike Wilson lines due to the appearance of
collinear singularities (indeed, the kinematic variables defined in
eq.~(\ref{cuspangle}) diverge in this limit). It is restored however,
upon dividing by the eikonal jets, thus linking the breakdown of scale
invariance in the soft function to the jet function, and hence to the
cusp anomalous dimension (see ref.~\cite{Gardi:2009qi}). These
considerations lead to a differential equation for the soft anomalous
dimension, whose \emph{minimal solution} up to three-loop order is the
so-called {\it dipole
  formula}~\cite{Gardi:2009qi,Becher:2009cu,Becher:2009qa}
\begin{equation}
\Gamma_n^{\rm dip.}\left(\{p_i\},\{ {\bf T}_i\},\mu,\alpha_s\right)
=-\frac12\hat{\gamma}_K(\alpha_s)\sum_{i<j}\log\left(\frac{-s_{ij}-i0}
{\mu^2}\right){\bf T}_i\cdot {\bf T}_j
+\sum_{i=1}^n\gamma_{J_i}(\alpha_s).
\label{gamdip}
\end{equation}
Here $\hat{\gamma}_K$ is the cusp anomalous dimension, which is known
up to three-loop order in
QCD~\cite{Korchemsky:1985xj,Korchemsky:1987wg,Moch:2004pa,Grozin:2014hna},
with the Casimir of the representation of the Wilson lines scaled out;
$\gamma_{J_i}$ is an anomalous dimension associated with collinear
singularities, and is also known to three-loop order for both quark
and gluon jets~\cite{Moch:2005tm,Gehrmann:2010ue}. We have introduced
the Mandelstam invariants
\begin{equation}
-s_{ij}-i0=2|p_i\cdot p_j|e^{-i\pi\lambda_{ij}},
\label{mandies}
\end{equation}
where $\lambda_{ij}=1$ if partons $i$ and $j$ are both incoming or
both outgoing, and $\lambda_{ij}=0$ otherwise.

As refs.~\cite{Gardi:2009qi,Becher:2009cu,Becher:2009qa} make clear,
there are two possible sources of the correction to the dipole formula
beyond two-loop order. The first is related to the fact that
eq.~(\ref{gamdip}) contains the cusp anomalous dimension with all
colour dependence scaled out, thus assuming that {\it Casimir scaling}
holds to all orders. In fact, this breaks for the first time at
four-loop order due to the appearance of new colour structures, quartic Casimirs, as has very recently been shown explicitly in
ref.~\cite{Boels:2017skl}. This implies that the form of
eq.~(\ref{gamdip}) will have to be modified beyond three loops~\cite{Gardi:2009qi,Becher:2009cu,Becher:2009qa}. The
second source of corrections starts already at three loops, and
constitutes a homogeneous solution to the differential equation for
the soft anomalous dimension derived in ref.~\cite{Gardi:2009qi}. This
implies a kinematic dependence only through conformally invariant
cross ratios
\begin{equation}
\rho_{ijkl}\equiv\frac{(-s_{ij})(-s_{kl})}{(-s_{ik})(-s_{jl})}
=\frac{(\beta_i\cdot \beta_j)\,(\beta_k\cdot \beta_l)}
{(\beta_i\cdot \beta_k)\,(\beta_j\cdot\beta_l)},
\label{rhodef}
\end{equation}
such that, up to three-loop order, the complete soft anomalous
dimension assumes the form
\begin{equation}
\Gamma_n(\{p_i\},\{ {\bf T}_i\},\mu,\alpha_s)=
\Gamma_n^{\rm dip.}(\{p_i\},\{ {\bf T}_i\},
\mu,\alpha_s)+\Delta_n(\{\rho_{ijkl}\},\{ {\bf T}_i\},\alpha_s)\,,
\label{gamnfull}
\end{equation}
where the correction 
\begin{equation}
\Delta_n(\{\rho_{ijkl}\},\{ {\bf T}_i\},\alpha_s) =\left(\frac{\alpha_s}{4\pi}\right)^3 \Delta_n^{(3)}(\{\rho_{ijkl}\},\{ {\bf T}_i\})+{\cal O}(\alpha_s^4)
\end{equation}
begins at three-loop order. Whether or not it is nonzero at this order
remained conjectural for a number of
years~\cite{Dixon:2009ur,DelDuca:2011xm,DelDuca:2011ae,Ahrens:2012qz,Naculich:2013xa,Caron-Huot:2013fea}.
Recently, however, it was calculated for the first time in
ref.~\cite{Almelid:2015jia}. Before we quote its form, note that for
any given four particles $\{i,j,k,l\}$, there are potentially 24 cross
ratios. However, eq.~(\ref{rhodef}) implies
\begin{equation}
\rho_{ijkl}=\rho_{jilk}=\rho_{klij}=\rho_{lkji},
\label{rhotrans1}
\end{equation}
which reduces the number of cross ratios to 6. In fact, further relations such as
\begin{equation}
\rho_{ijkl}=\frac{1}{\rho_{ikjl}},\quad
  \rho_{ijlk}\rho_{ilkj}=\rho_{ijkl}
\label{rhotrans2}
\end{equation}
can be used to write all the cross ratios in terms of just 2
independent cross ratios, which can be taken to be $\{\rho_{ijkl},
\rho_{ilkj}\}$. However, we will keep the second set of relations
implicit in the rest of this section to make the symmetry of the
expressions more manifest. The explicit form of the three-loop
correction to the dipole formula can then be written
\begin{align}\label{eq:expected_Delta}
 \Delta_n^{(3)}\left(\left\{\rho_{ijkl}\right\},\{{\bf T}_i\}\right) &= 16\,f_{abe}f_{cde} 
\Big\{
-C\, \sum_{i=1}^n \ \sum_{\substack{{1\leq j<k\leq n}\\ j,k\neq i}}\left\{{\rm \bf T}_i^a,  {\rm \bf T}_i^d\right\}   {\rm \bf T}_j^b {\rm \bf T}_k^c \\
\nonumber&+{\sum_{1\leq i<j<k<l\leq n}}
\Big[
 {\rm \bf T}_i^a  {\rm \bf T}_j^b   {\rm \bf T}_k^c {\rm \bf T}_l^d   \, {\cal F}(\rho_{ikjl},\rho_{iljk}) 
+{\rm \bf T}_i^a  {\rm \bf T}_k^b {\rm \bf T}_j^c   {\rm \bf T}_l^d    
\, {\cal F}(\rho_{ijkl},\rho_{ilkj})\\
\nonumber&\qquad\qquad\qquad
+ {\rm \bf T}_i^a   {\rm \bf T}_l^b  {\rm \bf T}_j^c    {\rm \bf T}_k^d 
\, {\cal F}(\rho_{ijlk},\rho_{iklj}) \Big]       
\Big\}\,,
\end{align}
where 
\begin{equation}
C=\zeta_5+2\zeta_2\zeta_3,
\label{Cdef}
\end{equation}
and the explicit form of the function ${\cal F}$ can be given by
introducing variables $\{z_{ijkl},\bar{z}_{ijkl}\}$ satisfying
\begin{equation}
z_{ijkl}\bar{z}_{ijkl}=\rho_{ijkl},\quad
(1-z_{ijkl})(1-\bar{z}_{ijkl})=\rho_{ilkj}.
\label{zdef} 
\end{equation}
With these definitions, one has
\begin{equation}
{\cal F}(\rho_{ijkl},\rho_{ilkj})=F(1-z_{ijkl})
-F(z_{ijkl}),
\label{calFdef}
\end{equation}
where
\begin{equation}
F(z)={\cal L}_{10101}(z)+2\zeta_2[{\cal L}_{001}(z)
+{\cal L}_{100}(z)].
\label{Fdef}
\end{equation}
In the previous equations and throughout this paper we suppress the
dependence of functions on the variables
$\bar{z}_{ijkl}$.  The function ${\cal L}_w(z)$ (where $w$ is a word
composed of zeroes and ones) is a single-valued harmonic polylogarithm (SVHPL)~\cite{Brown:2004}. These are
special combinations of harmonic polylogarithms~\cite{Remiddi:1999ew}
that are free of discontinuities, and thus single-valued, in the
kinematic region where $\bar{z}$ is equal to the complex conjugate of
$z$. This region is a subset of the so-called Euclidean region, where
all Mandelstam invariants are spacelike with $s_{ij}<0$. Unitarity of
massless scattering amplitudes dictates that they can only have
singularities due to the vanishing of Mandelstam invariants, which do
not occur in the Euclidean region of fixed angle scattering. Thus,
single-valuedness of the function $F$ reflects directly the analytic
structure of the underlying amplitude.

Although eq.~(\ref{eq:expected_Delta}) was first written down when the
explicit result was presented in ref.~\cite{Almelid:2015jia}, we will
show in this paper that the exact functional form of the answer can be
deduced without a full calculation of the relevant Feynman
integrals. Our first step is to consider the nature of the
mathematical functions that the result can depend on.

\section{Wilson lines and the Riemann sphere}
\label{sec:HPL}

We start with a general analysis aimed at identifying the class of functions that depend on a set of $n$ four-velocities in a rescaling-invariant way.
The discussion in this section is independent of the number of loops, and we
specialise to the three-loop case in the next section.

One starts by considering the product of Wilson lines of
eq.~(\ref{Sdef}), whose vacuum expectation value dictates the form of
$\Delta_n$. Without loss of generality, we may choose all Wilson lines
in eq.~(\ref{Sdef}) to be timelike and future-pointing
($\beta_i^0>0$). The Wilson-line operator of eq.~(\ref{Phidef}) is
invariant under rescaling of its four-velocity,
$\beta_i\rightarrow\kappa_i\beta_i$. This allows us to fix the value
of $\beta_i^2$ for non-lightlike Wilson lines without affecting the correlator. In principle this value
can be different for each Wilson line. However, upon choosing a common
value, a correlator of $n$ Wilson lines is completely determined by a
set of $n$ points in three-dimensional hyperbolic space ${\mathbb
  H}^3$, namely the locus of all coordinates
\begin{displaymath}
\beta^\mu:\quad
 (\beta^0)^2-(\beta^1)^2-(\beta^2)^2 -(\beta^3)^2=R^2,\quad \beta^0>0,
\end{displaymath}
for some constant $R$.  As noted already in section~\ref{sec:review},
any vacuum expectation value of Wilson-line operators depends only on
the cusp angles between pairs of lines, via the normalised scalar products defined in
eq.~(\ref{cuspangle}).  In the hyperbolic space ${\mathbb H}^3$, this
translates into the requirement that the soft anomalous dimension
depends only on the geodesic distances between pairs of points, rather
than on their absolute position. Thus, a configuration of $n$ Wilson
lines is mapped to a set of $n$ points in ${\mathbb H}^3$, modulo any
symmetry transformations that leave the hyperbolic space
invariant. The latter space is known as ${\rm Conf}_n({\mathbb H}^3)$,
namely the set of configurations of $n$ points on hyperbolic
three-space. Given that Feynman integrals can often be cast in terms
of iterated integrals, one expects that they would appear also in this
case. However, to date there has been no systematic study of iterated
integrals on ${\rm Conf}_n({\mathbb H}^3)$, which means that we do not know a suitable set of basis functions for the soft anomalous dimension linking multiple massive partons.

The situation is simpler, however, for lightlike Wilson lines. In
order to approach the lightlike limit (which corresponds to taking
$R\rightarrow 0$ in the above coordinate system), it is convenient to
use an alternative coordinate system, exploiting the so-called upper
half space model of hyperbolic three-space, in which one identifies
\begin{equation}\label{eq:H3R2R}
{\mathbb H^3}\simeq {\mathbb R}^2\times {\mathbb R}_+, 
\end{equation}
with coordinates $(x,y,r)\in {\mathbb R}^2\times {\mathbb R}_+$. A
given Wilson-line four-velocity $\beta_i^\mu$ may then be parametrised as
\begin{equation}
\beta_i^0=1 +\frac{r_i^2+x_i^2+y_i^2}{4},\quad
\beta_i^1=x_i,\quad
\beta_i^2=y_i,\quad
\beta_i^3=1-\frac{r_i^2+x_i^2+y_i^2}{4}.
\label{betaparam}
\end{equation}
Each Wilson line satisfies $\beta_i^2=r_i^2$, but where different  Wilson lines have distinct values of $r_i$.
The lightlike
limit corresponds to $r_i\rightarrow 0$, such that the points in hyperbolic space move to the boundary. The latter has the topology of a sphere, which may be identified with the plane ${\mathbb{R}^2}$ of eq.~\eqref{eq:H3R2R} upon stereographic projection. In the coordinate system of eq.~\eqref{eq:H3R2R},
products of four-velocities take the form
\begin{equation}
\beta_i\cdot \beta_j=\frac{1}{2}
\left[(x_i-x_j)^2+(y_i-y_j)^2+r_i^2+r_j^2\right]=\frac{1}{2}
\left(|z_i-z_j|^2+r_i^2+r_j^2\right),
\label{betaibetaj}
\end{equation}
where we have identified the boundary of $\mathbb{H}^3$ with the Riemann sphere and defined the complex coordinates
\begin{equation}
z_j=x_j+iy_j \,\,\,\,\text{and}\,\,\,\, \bar{z}_j=x_j-iy_j\,.
\label{zidef}
\end{equation}
In the lightlike limit \hbox{$r_i,r_j\rightarrow 0$}, conformally invariant
cross ratios are given by
\begin{equation}
\rho_{ijkl}=\frac{(\beta_i\cdot \beta_j)(\beta_k\cdot \beta_l)}
{(\beta_i\cdot \beta_k)(\beta_j\cdot \beta_l)}
\xrightarrow{r_i,r_j,r_k,r_l\rightarrow \,0}
|{z}_{ijkl}|^2\,=\, {z}_{ijkl} {\bar{z}}_{ijkl}\,,
\label{CICRz}
\end{equation}
with
\beq
{z}_{ijkl}\equiv\frac{z_{ij}\,z_{kl}}{z_{ik}\,z_{jl}},\quad
z_{ij}=z_i-z_j\,,
\eeq
and equivalent definitions for the complex conjugates.
Thus, cross ratios of scalar products of Wilson-line velocities map to squares of cross ratios of complex distances on the
Riemann sphere. As above
for ${\mathbb H}^3$, symmetries of the sphere, namely an ${\rm
  SL}(2,{\mathbb C})$ invariance, can be factored out because
 only the angles between the Wilson lines matter, so that only
 conformally invariant ratios appear. We may write the correspondence
implied by eq.~(\ref{CICRz}) more formally as
\begin{equation}
{\rm Conf}_n(\partial{\mathbb H}^3)\simeq {\cal M}_{0,n},
\label{M0n}
\end{equation}
where ${\cal M}_{0,n}$ is the moduli space of Riemann spheres with $n$
marked points. The advantage of this latter identification is that the
nature of iterated integrals for this space is
completely known: they can always be expressed as linear combinations of products of multiple 
polylogarithms, and the coefficients of this linear combination are rational functions~\cite{Brown:2009qja}.

The ${\rm SL}(2,{\mathbb C})$ invariance may be used to
fix the positions of three of the $n$ points. Specifically considering
the quadruple of particles $\{i,j,k,l\}$,  one may
choose
\begin{equation}
z_i=z_{ijkl},\quad z_j=0,\quad z_k=\infty,\quad z_l=1,
\label{zchoice}
\end{equation}
so that the only nontrivial position $z_i$ on the Riemann sphere may
be identified with the variable of eqs.~(\ref{zdef}). We have thus
succeeded in furnishing the variables $z_{ijkl}$ with a geometric
interpretation. 

Recall that we have started from a configuration of timelike future-pointing Wilson lines in Minkowski space. In the context of loop integrals it is convenient to use Euclidean kinematics where all Lorentz invariants $\beta_i\cdot \beta_j$ are negative. The two pictures are related by analytic continuation. We note however that the phases acquired by the Lorentz invariants in this process cancel in conformally invariant cross ratios, such as those in eq.~(\ref{CICRz}). Therefore our parametrization of the cross ratios in terms of ${z}_{ijkl}$ and ${\bar{z}}_{ijkl}$ remains valid in the Euclidean region. In defining the functions we always refer to the Euclidean region where $\bar{z}_{ijkl}=z^*_{ijkl}$. One may then consider relaxing this condition, treating ${z}_{ijkl}$ and
 $\bar{z}_{ijkl}$ as independent variables related to the cross ratios via eq.~(\ref{CICRz}).

It is well known that scattering amplitudes are multivalued functions, and
the discontinuities across the branch cuts are related to the concept
of unitarity. As a consequence, the soft anomalous dimension cannot be
an arbitrary combination of multiple polylogarithms, but it is
constrained by unitarity. A convenient tool to study the analytic
properties of multiple polylogarithms is the symbol and, more generally, the
coproduct~\cite{Chen,Goncharov:1998kja,GoncharovGalois,Goncharov.A.B.:2009tja,Brown:2009qja,Duhr:2011zq,Brown:2011ik,Duhr:2012fh,Brown:Notes},
which maps a polylogarithm to a certain tensor product of
polylogarithms of lower weight. Discontinuities only act in the first
factor of the coproduct~\cite{Duhr:2012fh,Brown:Notes}, and so the
first entry in the coproduct can only have branch cuts dictated by
unitarity~\cite{Gaiotto:2011dt,Abreu:2015zaa}. A massless scattering
amplitude can only have branch points when a scalar product between
two external momenta vanishes or becomes infinite. Rescaling invariance then implies that
the soft anomalous dimension has branch cuts starting only at points
where a conformally invariant cross ratio vanishes or becomes infinite. The relation
between the cross ratios and the complex variables $z_{ijkl}$ in
eq.~\eqref{CICRz} then implies that the scattering amplitude in the
Euclidean region must be a single-valued function of this complex
variable~\cite{Dixon:2012yy,Chavez:2012kn,Schnetz:2013hqa,Drummond:2013nda,DelDuca:2013lma,DelDuca:2016lad,Broedel:2016kls,Schnetz:2016fhy,DelDuca:2017peo}.

Single-valued multiple polylogarithms have been
studied extensively in the literature. They can be expressed as linear
combinations of products of multiple polylogarithms and their complex
conjugates such that all branch cuts cancel. Moreover, they inherit many of
the properties of ordinary polylogarithms: they form a
shuffle algebra and satisfy the same holomorphic differential
equations and boundary conditions as their multi-valued
analogues. There are several ways to explicitly construct
single-valued polylogarithms, based on the Knizhnik-Zamolodchikov
equation~\cite{Brown:2004,BrownSVMPLs}, the coproduct and the action
of the motivic Galois group on multiple
polylogarithms~\cite{Brown:2013gia,Brown:Notes,DelDuca:2016lad} and
the existence of single-valued primitives of multiple
polylogarithms~\cite{Schnetz:2016fhy}.

Let us discuss how the singularities of an amplitude manifest
themselves in this language. A known property of ${\cal M}_{0,n}$ is
that singularities occur only when marked points $z_i$ and $z_j$
coincide. This corresponds to two Wilson lines becoming collinear,
which is the only case in which the soft anomalous dimension itself
becomes singular.

At three-loop order, one may irreducibly connect at most four Wilson
lines. From the above discussion (e.g. eq.~(\ref{zchoice})), this
implies that only one independent $z_{ijkl}$ variable will occur in
each term in $\Delta_n^{(3)}$. Thus, only the simplest instance of
single-valued polylogarithms will show up, namely the single-valued
version associated to harmonic polylogarithms (SVHPLs) with
singularities only for $z\in\{0,1,\infty\}$, which have been studied
in detail in ref.~\cite{Brown:2004}. For more general correlators
involving more than four Wilson lines beyond three-loop order,
single-valued \emph{multiple} polylogarithms will appear, depending on
more than one $z_{ijkl}$ variable, corresponding to the fact that the
complex dimension of ${\cal M}_{0,n}$ is $n-3$.
  
To summarise, we expect that any rescaling-invariant function of $n$
four-velocities can be expressed in terms of single-valued multiple
polylogarithms on ${\cal M}_{0,n}$.  More precisely, we expect to
obtain a linear combination of products of single-valued multiple
polylogarithms and multiple zeta values (MZVs), whose coefficients are
rational functions of the $z_{i}$ variables and their complex
conjugates with poles at most when two points coincide. At this point
a comment is in order: SVHPLs with argument $z=1$ evaluate (if
convergent) to special combinations of MZVs called \emph{single-valued
  MZVs} (SVMZVs)~\cite{Brown:2013gia}.  In particular, the
single-valued version of $\zeta_{2n}$ is zero~\cite{Brown:2013gia}. It
is therefore tempting to restrict the set of MZVs that can appear to
SVMZVs, which would in particular imply that no powers of $\pi$ could
appear in the final answer. As we will see shortly, this restriction
is incorrect.  Indeed, the argument why only single-valued multiple
polylogarithms can appear applies only to the kinematic-dependent function, and does not extend to constants.

Let us conclude this section by commenting on how the analysis presented here is connected to similar results in the literature.
Above we have used the particular coordinate transformation of
eq.~(\ref{betaparam}) in order to implement the map of
eq.~(\ref{M0n}). Similar arguments for reinterpreting Wilson
lines have been made before. Reference~\cite{Chien:2011wz}, for
example, uses both coordinate and conformal transformations to map
Wilson lines to static charges in Euclidean AdS$_3$ space, such that
they move to the boundary of this space upon becoming lightlike. The
boundary of this space is a two-sphere, which can be mapped to ${\cal
  M}_{0,n}$ similarly to
eq.~(\ref{M0n}). Reference~\cite{Cheung:2016iub} also considers
mapping four-dimensional momenta to a ``celestial sphere'' at
infinity, aiming to develop a holographic picture of Minkowski-space
amplitudes.

\section{Ansatz for $\Delta_n^{(3)}$}
\label{sec:constraints}

The considerations of the previous section were generic and independent of the number of loops.
In this section we restrict ourselves to three loops, and we present the most general ansatz for $\Delta_n^{(3)}$
in terms of SVHPLs based on symmetries. The ansatz will depend on a certain number of free parameters that cannot be fixed from 
symmetries alone. These will be fixed in subsequent sections using input from special kinematic limits.

\subsection{Colour structure of $\Delta_n^{(3)}$}
The soft anomalous dimension depends on the colour quantum numbers and
the four-velocities of the Wilson lines. Since at three loops at most
four Wilson lines can be irreducibly connected by gluons, each
term in $\Delta_n^{(3)}$ can involve at most four colour generators
${\bf T}_i$, $1\le i\le n$. The colour structures that enter the soft
anomalous dimension have been proven in ref.~\cite{Gardi:2013ita} to
correspond to graphs that remain completely connected when the Wilson
lines are removed. The full set of such \emph{connected colour factors}
at three-loop order has been classified, and this then allows us to write
the most general form that $\Delta_n^{(3)}$ can take:
\begin{align}
\begin{split}
\label{eq:Delta_general_color}
\Delta_n^{(3)}&= \sum_{\{ i,j,k,l \}} f_{abe}f_{cde}\, {\bf T}_i^a {\bf T}_j^b {\bf T}_k^c {\bf T}_l^d \,A_{ijkl}
+\sum_{\{ i,j,k \}}  f_{abe} f_{cde} \left\{{\bf T}_i^a,{\bf T}_i^d\right\} {\bf T}_j^b {\bf T}_k^c \,B_{ijk}\\
&+\sum_{\{ i,j,k \}}  f_{abc}{\bf T}_i^a{\bf T}_j^b{\bf T}_k^c \,C_{ijk}
+\sum_{\{ i,j \}}   f_{abe} f_{cde} \left\{{\bf T}_i^a,{\bf T}_i^d\right\} \left\{{\bf T}_j^b,{\bf T}_j^c\right\} \,D_{ij}
+\sum_{\{ i,j \}}  {\bf T}_i\cdot {\bf T}_j \,E_{ij}\ ,
\end{split}
\end{align}
where the sums run over all sets $\{ i,j,\ldots \}$ of distinct Wilson
lines, and the coefficients $A_{ijkl}$, $B_{ijk}$ etc. are functions
of the four-velocities of the Wilson lines entering each colour factor.

Equation~\eqref{eq:Delta_general_color}, however, is still largely
over-complete. First, it was shown in
ref.~\cite{Gardi:2009qi,Becher:2009cu,Becher:2009qa} that colour
tripoles of the form $f_{abc}{\bf T}_i^a{\bf T}_j^b{\bf T}_k^c$ are
absent at any loop order (a corresponding kinematic function would violate rescaling invariance) and so we must have $C_{ijk}=0$ in
eq.~\eqref{eq:Delta_general_color}. Second, it is important to keep in
mind that $\Delta_n^{(3)}$ is an operator in colour space acting on
the hard amplitude ${\cal H}_n$ in eq.~\eqref{softcolfac}. The hard
function is a colour singlet, which implies that it must be
annihilated by the sum of all colour charge operators,
\beq\label{eq:colour_conservation} \left(\sum_{i=1}^n{\bf
  T}_i^a\right){\cal H}_n = 0\,.  \eeq We emphasise that the sum of
all colour charge operators does not vanish in general, but only when
it acts on a colour-singlet state. In practise, this means that
eq.~\eqref{eq:colour_conservation} can only be applied after all
colour operators have been commuted all the way to the right of the
expression. For example, we have \beq \left(\sum_{i=1}^n{\bf
  T}_i^a\right){\bf T}_j^b\,{\cal H}_n = if_{abc}\,{\bf T}_j^c\,{\cal
  H}_n \neq 0\,.  \eeq

In ref.~\cite{AlmelidPhD,Gardi:2016ttq,longinprep} the role of colour
conservation in the context of $\Delta_n^{(3)}$ was analysed in
detail, and a basis of colour structures that are independent after
eq.~\eqref{eq:colour_conservation} was imposed was worked out. 
Upon using colour conservation to write
eq.~\eqref{eq:Delta_general_color} in that basis, one observes that
not all of the coefficients in eq.~\eqref{eq:Delta_general_color} are
independent.  In particular, we may choose
$D_{ij}=E_{ij}=0$, and rescaling invariance implies that $B_{ijk}$ is
then a constant independent of $i$, $j$ and $k$.

To summarise, we find that the colour structure of $\Delta_n^{(3)}$ is very constrained, and the most general ansatz for the colour structure of $\Delta_n^{(3)}$ is
\begin{align}
\begin{split}
\label{eq:ansatz_colour}
\Delta_n^{(3)}&= \sum_{\{i,j,k,l\}} f_{abe}f_{cde}\, {\bf T}_i^a {\bf T}_j^b {\bf T}_k^c {\bf T}_l^d \,A_{ijkl}(\rho_{ikjl},\rho_{iljk})\\
&\,-16\,C\,\sum_{i=1}^n \ \sum_{\substack{{1\leq j<k\leq n}\\ j,k\neq i}}  f_{abe} f_{cde} \left\{{\bf T}_i^a,{\bf T}_i^d\right\} {\bf T}_j^b {\bf T}_k^c\ ,
\end{split}
\end{align}
where we have rewritten the summation in the second line, and pulled out an overall numerical factor, for later convenience.
From section~\ref{sec:HPL} we know that the functions $A_{ijkl}$ can be expressed in terms of SVHPLs, but they cannot be constrained any further by analysing colour structure alone. The functional form is, however, heavily constrained on general grounds by symmetries, as we will discuss in the next section.

\subsection{Symmetries}
\label{sec:symmetries}

Much like the scattering amplitude itself, the soft anomalous dimension must respect a number of symmetries and identities. Most directly, given that the external particles have been replaced by Wilson lines, the soft anomalous dimension admits Bose symmetry --- that is, invariance under the simultaneous interchange of both the colour and kinematic indices associated with any two lines $i$ and $j$. The functions $A_{ijkl}$ in eq.~\eqref{eq:ansatz_colour} can therefore depend on the indices $\{i,j,k,l\}$ only through their kinematic arguments. 
We then immediately see that we can rewrite eq.~\eqref{eq:ansatz_colour} in the form of eq.~\eqref{eq:expected_Delta}, but where now the coefficient $C$ and the function ${\cal F}$ are regarded as undetermined.
In other words, the problem of pinning down $\Delta_n^{(3)}$ amounts to determining ${\cal F}$ and~$C$.

Note that Bose symmetry together with the antisymmetry of the structure constants in eq.~\eqref{eq:expected_Delta} also implies that ${\cal F}$ is antisymmetric in its two arguments:
\beq\label{eq:F=-F}
{\cal F}(\rho_{ijkl},\rho_{ilkj}) = - {\cal F}(\rho_{ilkj},\rho_{ijkl})\,.
\eeq
We see from eqs.~\eqref{CICRz} and (\ref{zchoice}) that both $\rho_{ijkl}$ and $\rho_{ilkj}$ can be written in terms of
$z_{ijkl}$ and~$\bar{z}_{ijkl}$:
\beq
\rho_{ijkl} = z_{ijkl} \,\bar{z}_{ijkl} \,{\rm~~and~~}\,\rho_{ilkj} = (1-z_{ijkl})  (1-\bar{z}_{ijkl})\,.
\label{rhoToz}
\eeq
Combined with eq.~\eqref{eq:F=-F}, this means that ${\cal F}$ can be recast in the form of eq.~\eqref{calFdef}, where once again the function $F$ is to be interpreted as undetermined.

In addition to Bose symmetry, the function $\Delta_n^{(3)}$ has an additional property when seen as a function of the variables $z_{ijkl}$ and $\bar{z}_{ijkl}$: it must be real in the Euclidean region, and in particular when $\bar{z}_{ijkl} = z_{ijkl}^*$. SVHPLs are real-analytic functions of their argument~\cite{Brown:2004}, and hence complex conjugation corresponds to complex-conjugating the argument. This implies that the function $F$ is invariant under the interchange $\bar{z}_{ijkl} \leftrightarrow z_{ijkl}$. This symmetry acts on SVHPLs through reversal of words~\cite{Brown:2004}, 
\begin{equation}
{\cal L}_w(\bar{z})={\cal L}_{\tilde{w}}(z)+\ldots.
\label{Linvert}
\end{equation}
where $\tilde{w}$ is the word obtained upon reversing the word $w$, and
the ellipsis denotes terms proportional to multiple zeta values that can be worked out if needed, but are irrelevant for the following.


\subsection{Constraints from $\mathcal{N}=4$ Super Yang-Mills}
\label{sec:transcend}
Aside from symmetries, the function $F$ and the constant $C$ are
constrained by additional properties.  One such property comes from
the observation that $\Delta_n^{(3)}$ is the same in QCD as it is in ${\cal N}=4$ Supersymmetric Yang-Mills (SYM) theory, since at this order contributions sensitive to the differing matter content in these
theories are entirely contained in the dipole contribution to the soft
anomalous dimension given in eq.~(\ref{gamdip})~\cite{Dixon:2009ur}.
This is advantageous because $L$-loop amplitudes in ${\cal N}=4$ SYM
are expected to have uniform \emph{transcendental weight} (or just
\emph{weight}) $2L$. Multiplicative factors of $\epsilon^{-m}$ in
dimensional regularization contribute a factor $m$ to this weight,
dictating that the weight of the remaining function should be
$2L-m$. The soft anomalous dimension, which is associated with a
single pole in $\epsilon$ in the amplitude, is thus expected to have uniform
transcendental weight five at three loops. Although this property
remains conjectural, it is obeyed by all previously calculated
amplitudes in ${\cal N}=4$ SYM theory, {\it cf.}, e.g.,
ref.~\cite{Bern:2005iz,DelDuca:2009au,DelDuca:2010zg,Goncharov:2010jf,Dixon:2011pw,Dixon:2011nj,Dixon:2013eka,Golden:2014xqf,Dixon:2014voa,Drummond:2014ffa,Dixon:2014iba,Dixon:2015iva,Caron-Huot:2016owq,Dixon:2016nkn,Henn:2016jdu}.

The transcendental weight of a multiple polylogarithm corresponds to the number of integrations appearing in the definition of the function (where these integrations are required to take a specific form, see ref.~\cite{Brown:2004,Remiddi:1999ew,Dixon:2012yy}). For example, rational factors have weight zero, while logarithms correspond to integrating once over the kernel $dx/x$ and so have weight one. In fact, logarithms are the only functions that can appear at weight one in the space of multiple polylogarithms. The Riemann zeta value $\zeta_n$ is also assigned weight $n$ since it appears 
at special values of weight~$n$ polylogarithms. The weight of a product of two multiple polylogarithms is the sum of their individual weights. The SVHPL ${\cal L}_w(z)$ is a linear combination of products of ordinary harmonic polylogarithms (HPLs) and their complex conjugates, such that each term in the sum has weight equal to the length $|w|$ of the word $w$. The weight of ${\cal L}_w(z)$ is therefore defined to be $|w|$. 

It is additionally believed that the soft anomalous dimension in \hbox{${\cal N}=4$} SYM
is a \emph{pure} function --- that is, a function without kinematic-dependent prefactors. This can be seen by considering the maximally-helicity-violating (MHV) amplitudes in this theory, which are not themselves pure functions but are expected to be linear combinations of pure functions dressed by simple rational prefactors~\cite{ArkaniHamed:2010gh,ArkaniHamed:2012nw,Bern:2014kca,Bern:2015ple}. For example, the on-shell four-point amplitude is a linear combination of pure functions multiplying the three prefactors $1/(st)$, $1/(su)$, and $1/(tu)$, each of which corresponds to a different tree-level channel~\cite{Henn:2016jdu}. However, these rational prefactors only contribute to the hard function $\mathcal{H}_4$ in the factorization scheme of eq.~(\ref{Andef}), implying that the remaining factor $Z_4$ must be a pure function of uniform weight. Since, moreover, the soft anomalous dimension matrix is independent of the helicity structure of the underlying hard scattering process, it must be endowed with this property more generally. We conclude that $\Delta_n^{(3)}$ is a pure function of uniform weight, whose only kinematic dependence appears in the SVHPLs themselves.

\subsection{An ansatz for $\Delta_n^{(3)}$}
We now combine all these ingredients and construct an ansatz for the function $F$ and 
constant~$C$, and thus for $\Delta_n^{(3)}$.
Relabelling indices and using the permutation properties of the colour factors, eq.~\eqref{eq:expected_Delta} can be recast in the form:
\begin{align}
\label{Delta3form}
\begin{split}
\Delta_n^{(3)}&=16 \hspace{-.5cm} {\sum_{1\leq i<j<k<l\leq n}} \hspace{-.5cm} {\bf T}_i^a {\bf T}_j^b {\bf T}_k^c {\bf T}_l^d \,\Bigg[
  f_{abe}f_{cde} \Big( F(z_{iljk}) - F(z_{ikjl}) \Big) \\
&+f_{ace}f_{bde} \Big( F(z_{ilkj}) - F(z_{ijkl}) \Big)
+f_{ade}f_{bce} \Big( F(z_{iklj}) - F(z_{ijlk}) \Big)\Bigg]
\\
&
- 16C \,\,\sum_{i=1}^n \sum_{\substack{{1\leq j<k\leq n}\\ \,j,k\neq i}}  f_{abe} f_{cde} \left\{{\bf T}_i^a,{\bf T}_i^d\right\} {\bf T}_j^b {\bf T}_k^c \ .
\end{split}
\end{align}
For a given set of four particles $\{i,j,k,l\}$, this formula contains
the variables $z_{ijkl}$ with six permutations of the indices. Actually, we
may rewrite the formula in terms of a {\it single} permutation
$z_{ijkl}$, based on the fact that the cross ratio transformations of
eqs.~(\ref{rhotrans1}) and~(\ref{rhotrans2}) imply, through
eq.~(\ref{zdef}), the following relations:
\begin{equation}
z_{ijkl}=\frac{1}{z_{ikjl}}=1-z_{ilkj}=\frac{z_{ijlk}}{z_{ijlk}-1}.
\label{zperms}
\end{equation}
Equation~(\ref{Delta3form}) then becomes
\begin{align}
\label{Delta3form2}
\Delta_n^{(3)}&=16 \hspace{-.5cm} {\sum_{1\leq i<j<k<l\leq n}} \hspace{-.5cm} {\bf T}_i^a {\bf T}_j^b {\bf T}_k^c {\bf T}_l^d \,\Bigg[
  f_{abe}f_{cde} \Big( F(1-1/z_{ijkl}) - F(1/z_{ijkl}) \Big) \\
\nonumber&+f_{ace}f_{bde} \Big( F(1-z_{ijkl}) - F(z_{ijkl}) \Big)
+f_{ade}f_{bce} \Big( F(1/(1-z_{ijkl})) - F(z_{ijkl}/(z_{ijkl}-1)) \Big)\Bigg]
\\
\nonumber&
- 16C \,\,\sum_{i=1}^n \sum_{\substack{{1\leq j<k\leq n}\\ \,j,k\neq i}}  f_{abe} f_{cde} \left\{{\bf T}_i^a,{\bf T}_i^d\right\} {\bf T}_j^b {\bf T}_k^c \ .
\end{align}

A general ansatz for $F(z)$ consists of 32 weight five SVHPLs, 8
weight three SVHPLs multiplied by $\zeta_2$, 4 weight two SVHPLs
multiplied by $\zeta_3$, 2 weight one SVHPLs multiplied by $\zeta_4$,
and finally a general linear combination of the 2 weight five
constants $\zeta_5$ and $\zeta_{2}\zeta_3$, all with rational
prefactors.\footnote{The space of MZVs contains only even powers of $\pi$, thus forbidding the appearance of weight four SVHPLs.} This gives 48 distinct terms. Similarly, a general ansatz
for the constant $C$ consists of a general linear combination of the
2 independent weight five MZVs, with rational prefactors. However, this
na\"ive ansatz is overly large for two reasons. First, because of the
invariance under the interchange $z_{ijkl}\leftrightarrow\bar{z}_{ijkl}$ and the way
this symmetry acts on SVHPLs via reversal of words (see
eq.~\eqref{Linvert}), only palindromic combinations of weight five
SVHPLs need to be considered.
%
%
Second, the function $F(z_{ijkl})$ only appears in our ansatz (\ref{Delta3form2}) via
the differences ({\it cf.} eq.~(\ref{eq:expected_Delta})):
\begin{align}
\label{Fdiffs}
\begin{split}
{\cal F}(\rho_{ikjl},\rho_{iljk})=&\,F(1-1/z_{ijkl})-F(1/z_{ijkl}),\\
{\cal F}(\rho_{ijkl},\rho_{ilkj})=&\,F(1-z_{ijkl})-F(z_{ijkl}),\\
{\cal F}(\rho_{ijlk},\rho_{iklj})=&\,F(1/(1-z_{ijkl}))-F(z_{ijkl}/(z_{ijkl}-1)).
\end{split}
\end{align}
We should therefore only consider the number of linearly independent
combinations of SVHPLs that show up in these expressions (a trivial example is that a constant term in $F$ immediately drops out when considering the differences in eq.~(\ref{Fdiffs})). 
In fact, we can go one step further, as the three terms in the square brackets of
eq.~(\ref{Delta3form}) are not linearly independent due to the fact that
the colour factors appearing there are related by the Jacobi identity
\begin{equation}
f_{ace}f_{bde}=f_{abe}f_{cde}+f_{ade}f_{bce}.
\label{Jacobi}
\end{equation}
One may choose to eliminate any of the three products of structure
constants inside the square brackets in eq.~(\ref{Delta3form2}), after which only two of the combinations
\begin{align}\label{Fcombs}
\begin{split}
F_1(z)&\equiv F(1-1/z)-F(1/z)+F(1-z)-F(z),\\
F_2(z)&\equiv F(1/z)-F(1-1/z)+F(1/(1-z))-F(z/(z-1)),\\
F_3(z)&\equiv F(z)-F(1-z)+F(z/(z-1))-F(1/(1-z))=-F_1(z)-F_2(z),
\end{split}
\end{align}
will remain. For example, eliminating $f_{abe}f_{cde}$ one gets
\begin{align}
\label{Delta3form3}
\Delta_n^{(3)}&=16 \hspace{-.5cm} {\sum_{1\leq i<j<k<l\leq n}} \hspace{-.5cm} {\bf T}_i^a {\bf T}_j^b {\bf T}_k^c {\bf T}_l^d \,\Bigg[ f_{ace}f_{bde} F_1(z_{ijkl})+f_{ade}f_{bce} F_2(z_{ijkl})\Bigg]
\\
\nonumber&
- 16C \,\,\sum_{i=1}^n \sum_{\substack{{1\leq j<k\leq n}\\ \,j,k\neq i}}  f_{abe} f_{cde} \left\{{\bf T}_i^a,{\bf T}_i^d\right\} {\bf T}_j^b {\bf T}_k^c \ .
\end{align}
We should thus restrict our attention to the linear
combinations of $F(z)$ given in eq.~(\ref{Fcombs}).
Note that these functions are related by the transformations
\begin{align}
F_2(z)=F_1\left(\frac{z}{z-1}\right),\quad F_3(z)=F_1(1-z),
\quad F_2(z)=F_3\left(\frac{1}{z}\right).
\label{Ftrans}
\end{align}
We see from eq.~(\ref{zperms}) that upon taking $z=z_{ijkl}$ in (\ref{Ftrans}) these three relations correspond directly to exchanging momenta
$p_i\leftrightarrow p_j$, $p_i\leftrightarrow p_k$, and
$p_i\leftrightarrow p_l$ respectively. 
Moreover, the functions in eq.~(\ref{Fcombs}) are each symmetric under one of the above permutations, namely
\begin{align}
F_3(z)=F_3\left(\frac{z}{z-1}\right),\quad F_2(z)=F_2(1-z),
\quad F_1(z)=F_1\left(\frac{1}{z}\right).
\label{Fantisymm}
\end{align}
The kinematic relation $-F_3(z)=F_1(z)+F_2(z)$, in conjunction with the Jacobi identity (\ref{Jacobi}), allows us to express $\Delta_n^{(3)}$ in many ways.\footnote{We note in passing the structure shared by the kinematic relation $-F_3(z)=F_1(z)+F_2(z)$ and the colour Jacobi identity (\ref{Jacobi}), which is reminiscent of the known colour-kinematics duality for loop integrands~\cite{Bern:2008qj}.} In particular, we may rewrite (\ref{Delta3form3}) in a way that distinguishes the components which are symmetric or antisymmetric with respect to a given pair of Wilson lines. For example, we obtain
\begin{align}
\label{Delta3form4}
\Delta_n^{(3)}&=- 16C \,\,\sum_{i=1}^n \sum_{\substack{{1\leq j<k\leq n}\\ \,j,k\neq i}}  f_{abe} f_{cde} \left\{{\bf T}_i^a,{\bf T}_i^d\right\} {\bf T}_j^b {\bf T}_k^c \ 
\\
\nonumber&
+8 \hspace{-.5cm} {\sum_{1\leq i<j<k<l\leq n}} \hspace{-.5cm} {\bf T}_i^a {\bf T}_j^b {\bf T}_k^c {\bf T}_l^d \,\Bigg[
   f_{abe}f_{cde}\Big( F_1(z_{ijkl})-F_2(z_{ijkl})\Big)  
-
 (f_{ace}f_{bde} +f_{ade}f_{bce}) F_3(z_{ijkl})\Bigg]\,,
\end{align}
where the first term in the last line is manifestly antisymmetric in both colour and kinematics under $i\leftrightarrow j$ interchange, while the second is symmetric in both under this transformation.  This clarifies the interpretation of the combinations of $F(z)$ defined in eq.~(\ref{Fcombs}).

Having now understood that Bose symmetry and the Jacobi identity imply that only the combinations of $F(z)$ appearing in eq.~(\ref{Fcombs}) can contribute to the soft anomalous dimension, we would like to write down an ansatz which is not over-complete.  
There is, however, a subtlety in doing so, namely there can be
cancellations between the terms on the right-hand side of the expressions in eq.~(\ref{Fcombs}) involving functions of different arguments, and which are not necessarily manifest until functional identities amongst SVHPLs are taken into
account. To deal with this, one may first convert all SVHPLs to
involve the {\it same} argument, before reducing to a minimal basis of
independent functions such as the Lyndon basis of ref.~\cite{Radford} using the aforementioned shuffle algebra. Then, 
combinations of SVHPLs that vanish in
eqs.~(\ref{Fcombs}) can be excluded from our ansatz. In combination with the requirement of invariance under the interchange $\bar{z} \leftrightarrow z$, this results in an ansatz for $F(z)$ involving only 11 parameters:
\begin{align}\label{ansatz1}
F(z)&=a_1{\cal L}_{00000}+a_2{\cal L}_{00100}+a_3{\cal L}_{10001}
+a_4{\cal L}_{10101}+a_5\left({\cal L}_{01001} + {\cal L}_{10010}\right)\nonumber \\
&\quad+a_6\left[{\cal L}_{00101} + {\cal L}_{10100} + 2 ( {\cal L}_{00011} 
+ {\cal L}_{11000})  \right]
+a_7\left[{\cal L}_{11010} + {\cal L}_{01011} + 3 ( {\cal L}_{00011} 
+ {\cal L}_{11000}) \right]\nonumber \\
&\quad+a_8\,\zeta_2{\cal L}_{000}
+a_9\,\zeta_2\left({\cal L}_{001}+{\cal L}_{100}\right)
+a_{10}\,\zeta_3\,{\cal L}_{00}+a_{11}\,\zeta_2^2\,{\cal L}_0,
\end{align}
where each $a_i\in{\mathbb Q}$ is an undetermined rational numerical
coefficient, and we have suppressed the dependence of the SVHPLs on their
argument, ${\cal L}_w \equiv {\cal L}_w(z)$. Writing our ansatz for the 
constant~$C$ explicitly as
\begin{equation}
C=a_{12}\,\zeta_5+a_{13}\,\zeta_2\,\zeta_3\,,
\label{ansatz2}
\end{equation}
we have a total of 13 undetermined rational coefficients. In other
words, a correct understanding of the symmetries of $\Delta_n^{(3)}$
and the space of functions it can depend on determines its value up to
only 13 rational coefficients. In order to determine these numerical
coefficients using only general constraints, we now turn to the Regge
and collinear limits.

\section{The Regge limit}
\label{sec:Reggelimit}

\subsection{The Regge limit of $\Delta_4^{(3)}$}
\label{sec:Regge}

The study of scattering amplitudes in the high-energy or {\it Regge
  limit} predates the use of QCD as a field theory of strong
interactions (see e.g. ref.~\cite{Eden:1966dnq} for a review). In the
case of $2\rightarrow 2$ scattering, one may parametrise an amplitude
according to the centre-of-mass energy $s$ and the momentum transfer
$t$. The Regge limit then corresponds to
\begin{equation}
s\gg -t,
\label{Regge}
\end{equation}
and is such that amplitudes become dominated by a power-like growth in
energy. In perturbation theory at leading power in $(-t)/s$, this
manifests itself as logarithmically enhanced terms
\begin{equation}
\alpha_s^p\,\log^{m} \left(\frac{s}{-t}\right),
\quad m\leq p
\label{Ldef}
\end{equation}
dressing the Born amplitude, that can be resummed to all orders in
perturbation theory. The link between the Regge limit and infrared
singularities was first explored in
ref.~\cite{Korchemskaya:1996je,Korchemskaya:1994qp}. Wilson lines
naturally occur in both contexts (see also
ref.~\cite{Balitsky:2001gj}) and the subject has been more recently
studied in
refs.~\cite{DelDuca:2011xm,DelDuca:2011ae,DelDuca:2013ara,DelDuca:2013dsa,DelDuca:2014cya,Caron-Huot:2013fea,Caron-Huot:2016tzz,Caron-Huot:2017fxr}.
The Regge limit was first used as a constraint on the soft anomalous dimension in refs.~\cite{DelDuca:2011xm,DelDuca:2011ae} where the dipole formula was shown to generate all infrared singularities at leading- and next-to-leading logarithmic accuracy in the high-energy limit in the real part of the amplitude. These papers also showed that the absence of super-leading logarithms with $m>p$ in eq.~(\ref{Ldef}) already provides a useful constraint on $\Delta_4^{(3)}$. Below we will see that this is indeed so, but we will be able to go further and provide powerful constraints on $\Delta_4^{(3)}$ using the state-of-the-art knowledge of high-energy logarithms 
deduced using rapidity evolution equations \cite{Caron-Huot:2013fea,Caron-Huot:2017fxr}. Reference~\cite{Caron-Huot:2017fxr}
in particular established the structure of high-energy logarithms at three loops and examined the infrared singular structure of the Regge limit in detail, and we will make direct use of the conclusions reached there.

For the present analysis, we will use the information that at three
loops the dipole form of the soft anomalous dimension,
eq.~(\ref{gamdip}), correctly predicts the highest three powers of
large logarithms in the real part of the
amplitude~\cite{Caron-Huot:2017fxr}, and the highest two powers of
large logarithms in the imaginary part~\cite{Caron-Huot:2013fea}.
These statements can be translated into the constraint that these or
higher powers of large logarithms are absent from $\Delta^{(3)}_4$ in
all Regge limits. More specifically, the appearance of $\log^m
(s/(-t))$ is excluded for $m \ge 1$ in the real part and $m \ge 2$ in
the imaginary part of $\Delta^{(3)}_4$ in these limits. There is a
different Regge limit corresponding to every partition of the partons
$\{1,2,3,4\}$ into an incoming pair and outgoing pair. The Regge limit
in which partons $i$ and $j$ are incoming should be evaluated in the
region where the Mandelstam invariants $s\equiv s_{ij}$ and $s_{kl}$
with $\{i,j,k,l\}=\{1,2,3,4\}$ are positive while all others are
negative, as dictated by eq.~\eqref{mandies}. Conversely, our ansatz
is most naturally formulated in the Euclidean region where all
Mandelstam invariants are negative, since it is only here that
$\Delta_{4}^{(3)}$ is single-valued. Thus, to get to the Regge limit
in which partons $i$ and $j$ scatter into partons $k$ and $l$, we must
first analytically continue our ansatz into the appropriate Minkowski
region and then take the limit $s\gg (-t)$ for this set of incoming and
outgoing particles. This procedure will be described in more detail in
the remainder of this section.

\subsection{Constraints from the Regge limit}
\label{sec:Reggeconstrain}

Analytic continuation of invariants away from the Euclidean region
involves several steps that will be described in the following.

\paragraph{A. Analytic continuation to incoming Wilson lines.}
Let us start with our ansatz in the Euclidean region where all Mandelstam invariants are negative in the case of $n=4$ coloured partons, and any number of additional colourless particles, i.e., the momenta of the four partons do not sum to zero. We want to analytically continue our ansatz to the physical region where $\beta_1$ and $\beta_2$ are incoming while the other two Wilson lines are outgoing. From eq.~\eqref{mandies} we know that the phase of a Mandelstam invariant $s_{ij}$ is determined by whether both $\beta_i$ and $\beta_j$ are ingoing or outgoing or not. This in turn determines the phase of the cross ratios:
\begin{align}
\begin{split}
\rho_{1234}&\, = \frac{(-s_{12}-i0)(-s_{34}-i0)}{(-s_{13}-i0)(-s_{24}-i0)} = \left|\frac{s_{12}s_{34}}{s_{13}s_{24}}\right|\, e^{-2i \pi}\,, \label{regge_zzbar} \\ 
\rho_{1432} &\,= \frac{(-s_{14}-i0)(-s_{23}-i0)}{(-s_{13}-i0)(-s_{24}-i0)} = \left|\frac{s_{14}s_{23}}{s_{13}s_{24}}\right|\,. 
\end{split}
\end{align}
Although the cross ratios do not change their value, it would be incorrect to set the phase on the right-hand side of this equation to 1, because the amplitude has branch cuts. It is therefore important to keep track of such phases to ensure that the amplitude is evaluated on the correct Riemann sheet, starting from the Euclidean region where the amplitude is real. 
Phases are generated for all invariants that become timelike (in our case $s_{12}$ and $s_{34}$) but not for those that remain spacelike (all others).
In performing this analytic continuation we may choose paths for $s_{12}$ and $s_{34}$  along a loop in the upper half plane, consistent with the Feynman $+i0$ prescription ({\it cf.} eq.~(\ref{mandies})):
\beq\begin{split}
-s_{12}-i0\to  |s_{12}|\,e^{- i\pi t} {\rm~~and~~} -s_{34}-i0\to |s_{34}|\,e^{- i\pi t}\,,\textrm{ with } t\in [0,1)\,,
\end{split}\eeq
where $t=0$ corresponds to the starting point, the Euclidean region, while $t\to 1^-$ corresponds to the final value where the invariants are timelike.
Consequently, the cross ratio $\rho_{1234}$ moves along a circle
around the origin, in agreement with eq.~\eqref{regge_zzbar}: 
\beq
\rho_{1234}(t) = |\rho_{1234}|\, e^{-2i \pi t}\,,\textrm{ with } t\in [0,1)\,.  
\eeq
$\Delta_n^{(3)}$, however, depends on the cross ratios and the
Mandelstam variables only through the variables $(z,\bar z)\equiv
(z_{1234},\bar z_{1234})$ defined in eq.~\eqref{CICRz}, 
and so we need to work out the path travelled by $(z,\bar z)$ as $\rho_{1234}$ moves
along the circle. We start by inverting eq.~\eqref{rhoToz}, and we
find: \beq z = Z_+(\rho_{1234},\rho_{1432}) {\rm~~and~~} \bar{z} =
Z_-(\rho_{1234},\rho_{1432})\,, \eeq with \beq Z_{\pm} =
\frac{1}{2}\left(1+\rho_{1234}-\rho_{1432}
\pm\sqrt{(1-\rho_{1234}-\rho_{1432})^2-4\rho_{1234}\rho_{1432}}\right)\,.
\eeq We then see that $(z,\bar z)$ move along the paths parametrised
by \beq z(t) = Z_+(\rho_{1234}(t),\rho_{1432}) {\rm~~and~~} \bar{z}(t)
= Z_-(\rho_{1234}(t),\rho_{1432})\,.  \eeq 
The paths are shown in
fig.~\ref{fig:anal_cont}.
We see that, although we
started from a point where $\bar{z}=z^*$, this relation is no longer
valid for arbitrary $t$. As a consequence, $\Delta_4^{(3)}$ will not
be single-valued for generic $t$, and so the function may develop an
imaginary part and will not be single-valued after analytic
continuation. Note that the variables $z$ and $\bar z$ exchange
their roles at the end of the analytic continuation, such that $|z(0)|^2 =
|z(1)|^2$, in agreement with the observation that the absolute value
of $\rho_{1234}$ does not change. We also see from
fig.~\ref{fig:anal_cont} that $\bar{z}(t)$ crosses the branch cut
starting at $\bar{z}=0$ clockwise, while ${z}(t)$ avoids all branch
cuts. This is equivalent to the combined trajectory drawn by $\bar{z}(t)$ and $z(t)$ encircling the branch point at the origin.
The combined phases are given by \beq\begin{split} \lim_{t\to
  1}z(t)\bar{z}(t) &\,= |z|^2\,e^{-2i\pi}\,,\\ \lim_{t\to
  1}(1-z(t))(1-\bar{z}(t)) &\,= |1-z|^2\,,
\end{split}\eeq
in agreement with eq.~\eqref{regge_zzbar}. The previous equation is in fact sufficient to perform the analytic continuation of the SVHPLs that appear in our ansatz in eq.~\eqref{regge_zzbar}. Indeed, since only the branch cut starting at $z=0$ is crossed, we can use the shuffle algebra properties of SVHPLs to make all discontinuities explicit: ${\cal L}_w(z)$ has a branch point at $z=0$ if and only if the rightmost letter in the word $w$ is a `0'. We can thus use the shuffle algebra to recast ${\cal L}_w(z)$ in a form where the only words whose rightmost entry is a `0' are those of the form $(0,\ldots,0)$. These SVHPLs are just powers of logarithms of $z$ and $\bar{z}$ and are simple to analytically continue, e.g.,
\beq
{\cal L}_{10}(z) = {\cal L}_{0}(z)\,{\cal L}_{1}(z)-{\cal L}_{01}(z) \longrightarrow ({\cal L}_{0}(z)-2i\pi)\,{\cal L}_{1}(z)-{\cal L}_{01}(z)\,.
\eeq
Applying this procedure to every SVHPL in eq.~\eqref{ansatz1}, we can analytically continue our ansatz to the physical region where the Wilson lines 1 and 2 are incoming.

\begin{figure}[!t]
\begin{center}
\includegraphics[scale=0.3]{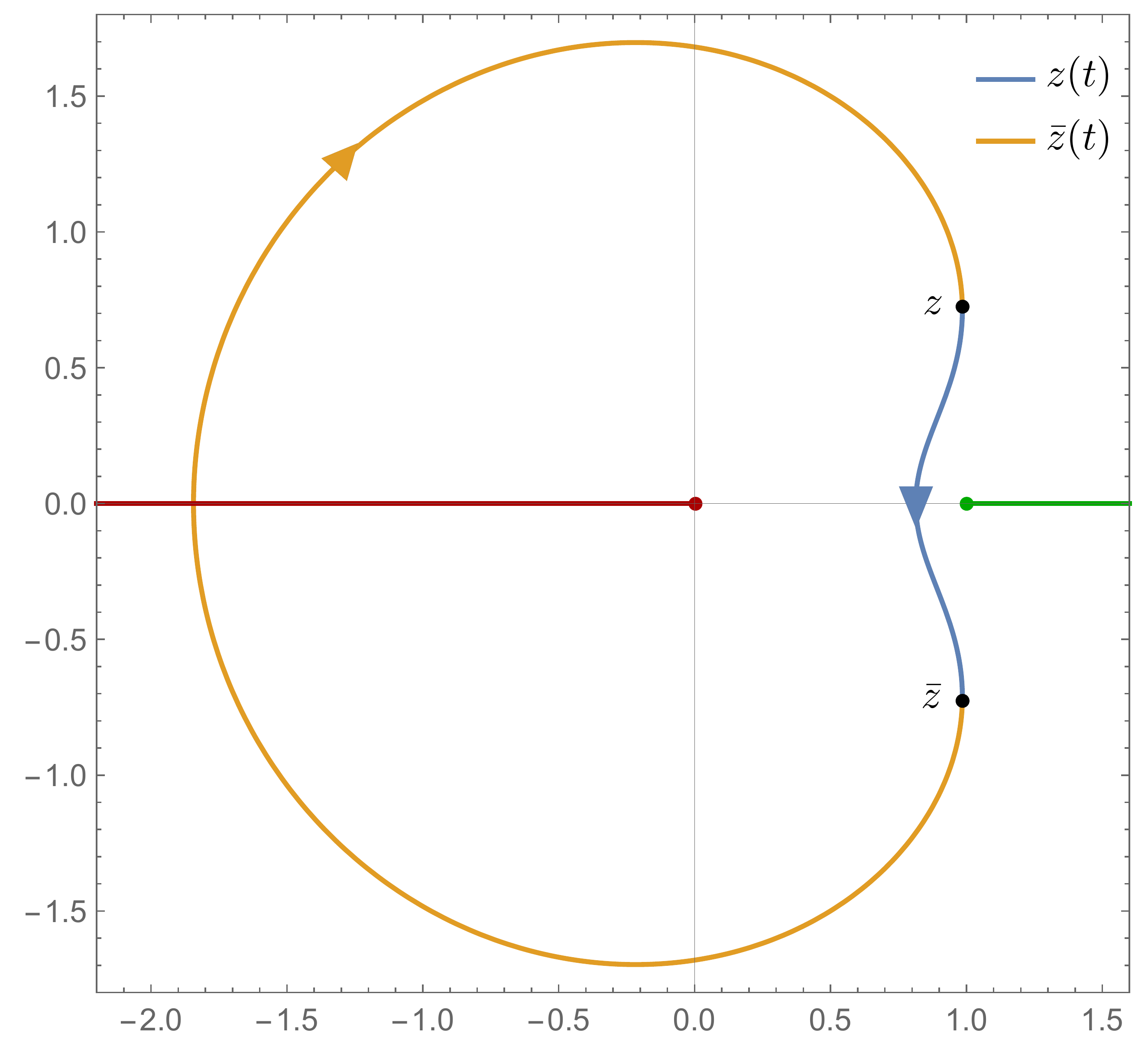}
\caption{\label{fig:anal_cont}Paths in the complex plane for the analytic continuation of the variables $z$ and $\bar{z}$ to the physical region where partons 1 and 2 are incoming.}
\end{center}
\end{figure}

Let us conclude by discussing how the analytic continuation changes if other Wilson lines are incoming. 
\begin{itemize}
\item If the Wilson lines 1 and 4 (or 2 and 3) are incoming, then only $\rho_{1432}$ acquires a phase:
\beq
(\rho_{1234},\rho_{1432}) \longrightarrow (\rho_{1234},\rho_{1432}\,e^{-2i\pi})\,.
\label{con2}
\eeq
Repeating the previous analysis, one finds that $\bar{z}(t)$ avoids all branch cuts, while ${z}(t)$ crosses the branch cut starting at ${z}=1$ in the clockwise direction. The imaginary part can be determined in a way similar to the previous case, by using the shuffle algebra to make all discontinuities at ${z}=1$ explicit:
\beq
{\cal L}_{01}(z) = {\cal L}_{0}(z)\,{\cal L}_{1}(z)-{\cal L}_{10}(z) \longrightarrow {\cal L}_{0}(z)\,({\cal L}_{1}(z)+2i\pi)-{\cal L}_{10}(z)\,.
\eeq
\item If the Wilson lines 1 and 3 (or 2 and 4) are incoming, then both $\rho_{1234}$ and $\rho_{1432}$ acquire a phase:
\beq
(\rho_{1234},\rho_{1432}) \longrightarrow (\rho_{1234}\,e^{2i\pi},\rho_{1432}\,e^{2i\pi})\,.
\label{con3}
\eeq
This time both $z(t)$ and $\bar{z}(t)$ cross branch cuts
starting at $z=0$ and $\bar{z}=1$, respectively, going
counterclockwise. This is equivalent to the contour drawn by $z(t)$ and $\bar{z}(t)$ together encircling the branch point at infinity in the clockwise direction, and we can extract the imaginary parts as in the previous cases. 
\end{itemize}
We have thus seen that analytic continuation to the physical region of $2$-to-$2$ scattering takes the function away from the region where it is single-valued, generating imaginary parts. Thus after analytic continuation the function will be expressed in terms of ordinary harmonic polylogarithms. We have also seen that for each pair of incoming particles, there is a distinct branch point -- one of the three $\left\{0,1,\infty\right\}$ -- which is encircled by the combined trajectory of $z(t)$ and~$\bar{z}(t)$.   

\paragraph{B. The momentum conserving limit.}
Having performed the analytic continuation, our ansatz is now valid in the region where two specific partons are incoming. Below we only discuss the case where the partons 1 and 2 are incoming, and all other cases are similar. The kinematics does not yet correspond to a massless 2-to-2 scattering, because this requires momentum conservation among the partonic momenta, in addition to the constraint $s_{12}+s_{13}+s_{23} =0$, with $s_{12}>0$ and $s_{23}<0$. Since we will be interested in the Regge limit $s_{12}\gg (-s_{23})$, we can assume without loss of generality that we work in a region where $s_{12}$ is greater than $(-s_{23})$. Imposing these constraints, we see that the cross ratios become
\beq
\rho_{1234} = \left(\frac{s_{12}}{s_{12}+s_{23}}\right)^2 {\rm~~and~~} \rho_{1432} = \left(\frac{s_{23}}{s_{12}+s_{23}}\right)^2 \,.
\eeq
It is easy to check that in the momentum conserving limit eq.~\eqref{rhoToz} implies:
\beq\label{eq:z_momentum_conserving}
\bar{z}= z = \frac{s_{12}}{s_{12}+s_{23}} > 1\,.
\eeq
Having started from a complex conjugate pair, $\bar{z}=z^*$, the momentum conserving limit corresponds to the situation where $z$ and $\bar{z}$ approach the real axis. Care is needed, however, because $z$ and $\bar{z}$ approach the real axis from opposite sides. Indeed, if we assume that in the Euclidean region $z$ and $\bar{z}$ were in the upper and lower half planes respectively, then after analytic continuation $z$ and $\bar{z}$ have negative and positive imaginary parts, respectively. Hence, in the momentum conserving limit $z$ approaches the real axis from below, while $\bar{z}$ approaches it from above. Equation~\eqref{eq:z_momentum_conserving} then implies that some harmonic polylogarithms may develop opposite imaginary parts in the limit, e.g.,
\begin{align}
\log (1-z) &\xrightarrow{s_{12}+s_{13}+s_{23} =0} \log \left(\frac{-s_{23}}{s_{12}+s_{23}} \right) + i \pi\,,\\
\log (1- \bar z) &\xrightarrow{s_{12}+s_{13}+s_{23} =0}  \log \left(\frac{-s_{23}}{s_{12}+s_{23}}\right) - i \pi\,.
\end{align} 

Let us conclude by commenting on the class of functions that appear in the momentum conserving limit. Since $\bar{z}=z$, we can write $\Delta_{4}^{(3)}$ entirely in terms  of ordinary HPLs in the single variable $z/(z-1)=s_{12}/(-s_{23})$, in agreement with all known results for on-shell four-point amplitudes in QCD and ${\cal N}=4$ SYM~\cite{Anastasiou:2000kg,Anastasiou:2000ue,Anastasiou:2001sv,Glover:2001af,Bern:2002tk,DeFreitas:2004kmi,Bern:2005iz,Bern:2006ew,Naculich:2008ew,Naculich:2008ys,Naculich:2009cv,Henn:2016jdu}.

\paragraph{C. Constraints from the Regge limit.}
Having at our disposal the ansatz in the physical region of a 2-to-2 scattering, we can proceed and consider its Regge limit.
There are three different choices for the incoming particles, and for each choice we can consider two different Regge limits, corresponding to forward and backward scattering. In the following we discuss one of these limits in detail, and we only comment on the other limits at the very end. 

Let us consider the Regge limit of the scattering where the partons 1 and 2 are incoming and $s_{12} \gg (-s_{23})$. We know that our ansatz can be written in terms of HPLs in $x\equiv s_{12}/(-s_{23})$, and we can expand each HPL in a power series in $1/x\ll 1$. Dropping power-suppressed terms, we find that the functions $F_i$ in eq.~\eqref{Fcombs} reduce to
\begingroup
\allowdisplaybreaks
\begin{align}\label{Reggelim1}
{\rm Re}(F_1)&\xrightarrow{s_{12} \gg -s_{23}} 
-\frac{8a_1}{15} L^5 
+ \left(16 a_1 - 12 a_2 + 16 a_6 + 24 a_7 - \frac{8 a_8}{3}\right) 
  \zeta_2 L^3 + (4 a_{10}+ 16 a_7) 
 \zeta_3  L^2 \nonumber \\
&+ 
   (-48 a_1 - 4 a_{11} + 48 a_3 - 48 a_4 + 48 a_5 -288a_6-432a_7+ 24 a_8 + 24 a_9) \zeta_2^2 L\nonumber \\ 
& +  (12 a_{10} + 48 a_2 - 24 a_3 + 48 a_4 + 24 a_5 - 24 a_6 + 72 a_7 - 
    8 a_9) \zeta_2 \zeta_3 \nonumber \\
&+(-24a_2 + 12a_3 - 8a_4 - 36a_5 + 12a_6 - 4a_7)\zeta_5,\nonumber \\
{\rm Re}(F_2)&\xrightarrow{s_{12} \gg -s_{23}}  \frac{4a_1}{15} L^5 
+\left(8a_2+12a_3+12a_5+20a_6+28a_7+\frac{4a_8}{3}  \right)
\zeta_2 L^3\nonumber \\
&+(-2a_{10}-8a_7)\zeta_3 L^2
+(2a_{11}+48a_2-24a_3+48a_4+72a_5+168a_6+264a_7-24a_9)\zeta_2^2 L\nonumber \\
&+(12a_{10}-24a_2+24a_3-48a_4-24a_5+24a_6+72a_7+4a_9)\zeta_2\zeta_3\nonumber \\
&+(12 a_2-6a_3+4 a_4 +18a_5-6 a_6 +2 a_7)\zeta_5,\nonumber \\
{\rm Re}(F_3)&\xrightarrow{s_{12} \gg -s_{23}}  \frac{4a_1}{15}L^5
+\left(-16a_1+4a_2-12a_3-12a_5-36a_6-52a_7+\frac{4a_8}{3}\right)\zeta_2 L^3\nonumber \\
&+(-2a_{10}-8a_7)\zeta_3 L^2
+(48a_1+2a_{11}-48a_2-24a_3-120a_5+120a_6+168a_7-24a_8)\zeta_2^2 L\nonumber \\
&+(-24a_{10}-24a_2-144a_7+4a_9)\zeta_2\zeta_3\nonumber \\
&+(12a_2-6a_3+4a_4+18a_5-6a_6+2a_7)\zeta_5, 
\end{align}
and
\begin{align}\label{Reggelim2}
\frac{1}{\pi}{\rm Im}(F_1)&\xrightarrow{s_{12} \gg -s_{23}} \frac{4a_1}{3} L^4
+(-16a_1+18a_2-24a_6-36a_7+4a_8)\zeta_2L^2+(-4a_{10}-16a_7)\zeta_3 L\nonumber \\
&+\left(\frac{48a_1}{5}+2a_{11}+18a_2-24a_3+24a_4-24a_5+120a_6+180a_7
-8a_8-12a_9\right)\zeta_2^2,\nonumber \\
\frac{1}{\pi}{\rm Im}(F_2)&\xrightarrow{s_{12} \gg -s_{23}} 
\left(\frac{a_2}{6}+a_3+a_5+\frac{7a_6}{3}+\frac{10a_7}{3}\right)L^4\nonumber \\
&+(-2a_2-6a_3+4a_4+10a_5+2a_6+6a_7-2a_9)\zeta_2 L^2\nonumber \\
&+(8a_{10}+4a_3-8a_4-4a_5+4a_6+44a_7)\zeta_3 L\nonumber \\
&+\left(\frac{48a_1}{5}+2a_{11}+\frac{18a_2}{5}-68a_3+\frac{64a_4}{5}+12a_5
+76a_6+\frac{632a_7}{5}-8a_8-24a_9\right)\zeta_2^2,
  \nonumber \\
\frac{1}{\pi}{\rm Im}(F_3)&\xrightarrow{s_{12} \gg -s_{23}} 
\left(-\frac{4a_1}{3}-\frac{a_2}{6}-a_3-a_5-\frac{7a_6}{3}-\frac{10a_7}{3}
\right)L^4\nonumber \\
&+(16a_1-16a_2+6a_3-4a_4-10a_5+22a_6+30a_7-4a_8+2a_9)\zeta_2 L^2 \\
&+(-4a_{10}-4a_3+8a_4+4a_5-4a_6-28a_7)\zeta_3 L\nonumber \\
&+\left(12a_5-\frac{96a_1}{5}-4a_{11}-\frac{108a_2}{5}+92a_3-\frac{184a_4}{5} -196a_6+\frac{1532a_7}{5}+16a_8+36a_9\right)\zeta_2^2, \nonumber
\end{align}
\endgroup
where we have adopted the notation $L \equiv \log x$. 
These expressions can be compared to the results of ref.~\cite{Caron-Huot:2017fxr,Caron-Huot:2013fea}, as mentioned above. 
This requires the coefficients of  $L^{m}$, with $m\geq 1$ in the real part of the amplitude and $m\geq 2$ in the  
imaginary part, to vanish in eqs.~\eqref{Reggelim1} and \eqref{Reggelim2}. We then find six independent conditions on the undetermined coefficients $a_i$ in our ansatz for $F(z)$.

We have carried out the same analysis in the remaining five Regge limits, and the
expansion of our ansatz in two of these limits -- where partons 1 and 3 are incoming and $s_{13}\gg (-s_{14})$, and where 
partons 1 and 4 are incoming and $s_{14}\gg (-s_{13})$ --
 are presented in Appendix~\ref{app:regges}. Each Regge limit gives rise to six constraints, but those coming from limits involving the same pairs of incoming (or outgoing) partons are identical. Any pair of Regge limits involving different incoming (or outgoing) partons gives rise to eight independent constraints between them, after which considering additional Regge limits doesn't give rise to further constraints.
Putting together the constraints from the above Regge limit and one of the limits considered in Appendix~\ref{app:regges}, we can thus fix 8 out of the 13 free
parameters in our ansatz in eq.~\eqref{ansatz1}:
\begin{align}\label{aconditions_regge}
\begin{split}
\!\!\!\!(a_1,\ldots,a_8) &= \left(0,\frac{a_{10}}{10},-\frac{a_{10}}{10}-\frac{a_{11}}{48},\frac{a_9}{2}-\frac{3a_{10}}{20}-\frac{a_{11}}{12},\frac{a_{10}}{10}+\frac{a_{11}}{48},\frac{7a_{10}}{20},-\frac{a_{10}}{4},-\frac{3a_{10}}{5}\right).
\end{split}
\end{align}
Having discussed the Regge limit in detail, we now turn to the kinematic limit in which two Wilson lines become collinear.

\section{The collinear limit}
\label{sec:collinearlimit}

\subsection{Collinear limits}
\label{sec:collinear}

As has already been considered in
refs.~\cite{Becher:2009qa,Dixon:2009ur}, further kinematic constraints
on the soft anomalous dimension arise from collinear limits. We briefly review this argument here.
Let us now work in the kinematic region where all $n$ coloured particles carrying momenta $\{p_k\}$ are outgoing, and consider the limit in which two of these partons, $i$ and $j$, become collinear.  In this limit $p_i\cdot p_j\rightarrow
0$, resulting in kinematic divergences inversely proportional to $p_i \cdot p_j$. It is well known that for final-state collinear partons\footnote{In
  the case of a space-like splitting, factorisation is
  violated~\cite{Catani:2011st,Forshaw:2012bi}.} these divergences factorise~\cite{Bern:1999ry,Kosower:1999xi,Feige:2014wja,Catani:2003vu}, such that one may write
\begin{equation}
{\cal A}_n(p_1,p_2,\{p_j\};\mu,\epsilon,\alpha_s)
\xrightarrow{1\parallel 2}{\bf Sp}(p_1,p_2,{\bf T}_1,{\bf T}_2;\mu,\epsilon,\alpha_s)\,
{\cal A}_{n-1}(P,\{p_j\},\mu,\epsilon,\alpha_s).
\label{colfac}
\end{equation}
Without loss of generality, we have taken particles 1 and 2 collinear,
where $\{p_j\}$, $j=3\ldots n$ denotes the set of remaining
momenta. The right-hand side contains the $(n-1)$-particle amplitude in
which the momenta $p_1$ and $p_2$ are replaced by the sum $P=p_1+p_2$,
multiplied by a universal {\it splitting function} ${\bf Sp}$, which collects
all singular contributions to the amplitude due to particles
1 and 2 becoming collinear. 
Care must be taken to interpret the colour structure of eq.~(\ref{colfac}). The amplitudes
${\cal A}_n$ and ${\cal A}_{n-1}$ live in $n$-parton and
$(n-1)$-parton colour space respectively. However, one may write a
colour generator on the line of momentum $P$ as
\begin{equation}
{\bf T}={\bf T}_1+{\bf T_2},
\label{Tdef}
\end{equation}
which promotes the amplitude ${\cal A}_{n-1}$ to live in $n$-parton
colour space after all. 
Crucially, the splitting amplitudes ${\bf Sp}$ must only depend on 
the quantum numbers of particles 1 and 2. It is this property that we wish to exploit, following refs.~\cite{Becher:2009qa,Dixon:2009ur}, in order to  
constrain the soft anomalous dimension. Let us briefly recall how the constraint arises, before analysing its implications regarding our ansatz.

One starts with the observation that infrared factorization, according to 
eq.~(\ref{Andef}), holds separately for the $n$ and $(n-1)$-parton\footnote{Note that ${\cal A}_{n-1}$ in eq.~(\ref{colfac}) is evaluated at $P^2=0$, and hence it obeys the usual 
 soft-collinear factorisation formula for massless partons scattering in eq.~(\ref{softcolfac}).} amplitudes in eq.~(\ref{colfac}):
\begin{subequations}
\label{fac2}
\begin{align}
{\cal A}_n(p_1,p_2,\{p_j\},\mu,\epsilon,\alpha_s)&=Z_n(p_1,p_2,\{p_j\},{\bf T}_1,{\bf T}_2,
\{{\bf T}_j\},\epsilon,\alpha_s(\mu_f^2))
\,{\cal H}_n(p_1,p_2,\{p_j\};\mu,\mu_f,\epsilon)\,,\label{Soft_fact_An}\\
{\cal A}_{n-1}(P,\{p_j\},\mu,\epsilon,\alpha_s)&=Z_{n-1}(P,\{p_j\},{\bf T},
\{{\bf T}_i\},\epsilon,\alpha_s(\mu_f^2))
\,{\cal H}_{n-1}(P,\{p_j\},\mu,\mu_f,\epsilon,\alpha_s)\,.\label{Soft_fact_An-1}
\end{align}
\end{subequations}
Kinematic divergences in the collinear limit $p_1 \cdot p_2\to 0$ appear in eq.~(\ref{Soft_fact_An}) in both $Z_n$ and the hard function ${\cal H}_n$.
By analogy with eq.~(\ref{colfac}), the hard function may be
factorised in the collinear limit according to
\begin{equation}
\!{\cal H}_n(p_1,p_2,\{p_j\};\mu,\mu_f,\epsilon,\alpha_s)
\xrightarrow{1\parallel 2}{\bf Sp}_{\cal H}(p_1,p_2,{\bf T}_1,{\bf T}_2;
\mu,\mu_f,\epsilon,\alpha_s)
{\cal H}_{n-1}(P,\{p_j\};\mu,\mu_f,\epsilon,\alpha_s),
\label{Hnfac}
\end{equation}
where ${\bf Sp}_{\cal H}$ is an appropriate splitting function
collecting all terms which are singular as $P^2\rightarrow 0$, such
that the $(n-1)$-particle hard function ${\cal H}_{n-1}$ may be
evaluated with $P^2=0$. Of course, all quantities in eq.~(\ref{Hnfac}) are infrared finite.

Equations~(\ref{fac2}) together with eqs.~(\ref{colfac}) and (\ref{Hnfac}) implies the condition
\begin{align}
\begin{split}
{\bf Sp}_{\cal H}(p_1,p_2,{\bf T}_1,{\bf T}_2,&\,\mu,\mu_f,\epsilon,\alpha_s)
=Z_n^{-1}(p_1,p_2,\{p_j\},{\bf T}_1,{\bf T}_2,\{{\bf T}_j\},\epsilon,\alpha_s(\mu_f^2))\\
&\quad \times {\bf Sp}(p_1,p_2,{\bf T}_1,{\bf T}_2,\mu,\epsilon,\alpha_s)\,\,
Z_{n-1}(P,\{p_j\},{\bf T},\{{\bf T}_j\},\epsilon,\alpha_s(\mu_f^2)),
\label{SpH}
\end{split}
\end{align}
where all quantities must be evaluated in the limit $P^2\to 0$. This equation implies a highly non-trivial cancellation
between both colour and kinematic dependence on the right-hand side,
such that the left-hand side depends only on the quantum numbers of
the two particles becoming collinear.

Given that the amplitude splitting function ${\bf Sp}$ does
not depend on the infrared factorisation scale $\mu_f$, one may
differentiate eq.~(\ref{SpH}) and use eq.~(\ref{rengroup}) to obtain
the condition
\begin{equation}\begin{split}
\frac{d}{d\ln\mu_f}&\,{\bf Sp}_{\cal H}(p_1,p_2,{\bf T}_1,{\bf T}_2,\mu,\mu_f,\epsilon,\alpha_s)\\
&\,=\Gamma_{\bf Sp}(p_1,p_2,{\bf T}_1,{\bf T}_2,\mu_f,\alpha_s(\mu_f^2))\,
{\bf Sp}_{\cal H}(p_1,p_2,{\bf T}_1,{\bf T}_2,\mu,
\mu_f,\epsilon,\alpha_s),
\label{SpHrengroup}
\end{split}
\end{equation}
where we have defined
\begin{equation}
\begin{split}
\Gamma_{\bf Sp}&(p_1,p_2,{\bf T}_1,{\bf T}_2,\mu_f,\alpha_s(\mu_f^2))\\
&\,\equiv \Gamma_n(p_1,p_2,\{p_j\},{\bf T}_1,{\bf T}_2,\{{\bf T}_j \},\mu_f,\alpha_s(\mu_f^2))
-\Gamma_{n-1}(P,\{p_j\},{\bf T},\{ {\bf T}_j \},\mu_f,\alpha_s(\mu_f^2))\,.
\label{GamSpdef}
\end{split}
\end{equation}
Given that the quantity on the left-hand side of this equation depends
manifestly on the quantum numbers of partons 1 and 2 only, this must
also be true for the right-hand side. Upon making the decomposition of
eq.~(\ref{gamnfull}) (valid up to three-loop order), one may further decompose
\begin{align}\label{GamSpdecomp}
\begin{split}
\Gamma_{\bf Sp}(p_1,p_2,{\bf T}_1,&\,{\bf T}_2,\mu_f,\alpha_s(\mu_f^2))=\Gamma^{\rm dip.}_{\bf
    Sp}(p_1,p_2,{\bf T}_1,{\bf T}_2,\mu_f,\alpha_s(\mu_f^2))\\
&\quad  +\Delta_n(\{\rho_{ijkl}\},{\bf T}_1,{\bf T}_2,\{ {\bf T}_j \},\alpha_s(\mu_f^2))-\Delta_{n-1}(\{\rho_{ijkl}\},{\bf T},\{{\bf T}_j \},\alpha_s(\mu_f^2)).
\end{split}
\end{align}
We want to take the
kinematic limit in which partons 1 and 2 become collinear. To this
end, we may parametrise each momentum according to:
\begin{equation}
p_1=xP+k,\quad p_2=(1-x)P-k,
\label{p1p2param}
\end{equation}
where $k$ is a small momentum to allow $P^2\neq 0$, while
$p_1^2=p_2^2=0$. 
The first term in eq.~(\ref{GamSpdecomp}) is then
found to be~\cite{Dixon:2009ur}
\begin{align}
\label{GamSpdipdef}
&\Gamma^{\rm dip.}_{\bf Sp}\,(p_1,p_2,{\bf T}_1,{\bf T}_2,\lambda,\alpha_s(\lambda^2))=
\gamma_{J_1}(\alpha_s(\lambda^2))+\gamma_{J_2}(\alpha_s(\lambda^2))
\,-\gamma_{J_P}(\alpha_s(\lambda^2))
\\
&-\frac12\hat{\gamma}_K
(\alpha_s(\lambda^2))\Big[\ln\Big(\frac{2|p_1\cdot p_2|e^{-i\pi\lambda_{12}}} {\lambda^2}\Big){\bf T}_1\cdot {\bf
    T}_2 \,-{\bf T}_1\cdot({\bf T}_1+{\bf T}_2)\ln x-{\bf T}_2\cdot
  ({\bf T}_1+{\bf T}_2)
  \ln(1-x)\Big]\,,\nonumber
\end{align}
in terms of the quantities appearing in eq.~(\ref{gamdip}). Note that here 
$\lambda_{12}=1$ as we assumed that $p_1$ and $p_2$ both belong to the final state.
Equation (\ref{GamSpdipdef}) 
by itself depends only on the quantum numbers of the two particles
becoming collinear, which finally leads to an important constraint on
the $\Delta_n$, namely that the difference
\begin{equation}
\Delta^{(3)}_{\bf Sp}({\bf T}_1,{\bf T}_2)=\left[
\Delta^{(3)}_n(\{\rho_{ijkl}\},{\bf T}_1,{\bf T}_2,\{{\bf T}_j\})
-\Delta^{(3)}_{n-1}(\{\rho_{ijkl}\},{\bf T},\{{\bf T}_j\})
\right]_{p_1\parallel p_2}
\label{delcoll}
\end{equation}
can only depend on the quantum numbers of the two particles
that are becoming collinear. Note that the right-hand side of eq.~\eqref{delcoll} is
evaluated in the limit where $p_1$ and $p_2$ have become collinear. 
As suggested by our notation, this quantity has no kinematic dependence. To see
this, first note that universality of the splitting function implies
that the result should be independent of the number of Wilson lines
$n$. When $n=2$, colour conservation implies that there is only one
independent colour structure (a single quadratic Casimir), such that
the dipole formula furnishes the complete splitting function, and the
correction term vanishes: $\Delta_2^{(3)}=0$. This in turn implies 
\begin{equation}\begin{split}
\label{DelSp3}
\Delta^{(3)}_{\bf Sp}({\bf T}_1,{\bf T}_2)&\,=\left[\Delta^{(3)}_3(-{\bf T}_1-{\bf T}_2,{\bf T}_1,{\bf T}_2)-\Delta_2^{(3)}({\bf T}_1,{\bf T}_2)
\right]_{p_1\parallel p_2}\\
&\,=\left.\Delta^{(3)}_3(-{\bf T}_1-{\bf T}_2,{\bf T}_1,{\bf T}_2)\right|_{p_1\parallel p_2}\,,
\end{split}
\end{equation}
but since there are no conformally invariant cross ratios that one can form
from three particle momenta, the right-hand side evaluates to a
constant. Universality of the splitting amplitude then implies that the 
same is true for all~$n$.


\subsection{Constraints from two-particle collinear limits}
\label{sec:collconstrain}

Equation~(\ref{delcoll}) dictates that when two partons become
collinear, the difference $\Delta_n^{(3)} - \Delta_{n-1}^{(3)}$ can
only depend on the colour and kinematic degrees of freedom of the two
partons becoming collinear. In fact, we saw that this difference must
be a constant, independent of the number of Wilson lines $n$ (provided
one takes $n\geq 3$, so that $\Delta_n^{(3)}\neq 0$). These properties
can be imposed as further constraints on our ansatz for $F(z)$. To do
so, we specialize to three- and four-parton scattering, and set $i =
1, j =2, k=3, l=4$. We do not impose momentum conservation (i.e. we
consider the situation in which any number of colour singlet particles
may carry additional momentum). This will allow us to consider pairs
of particles becoming collinear, without simultaneously restricting
the momenta of other coloured particles. We also choose to focus on
the collinear limit in which $p_1 \parallel p_2$. It will prove
convenient to define the basis of colour generators as follows~\cite{AlmelidPhD}:
\begin{align} 
 \label{color_basis}
\begin{split}
{\bf T}_A = {\bf T}_1 + {\bf T}_2,  &\qquad
{\bf T}_B = {\bf T}_1 - {\bf T}_2,  \\
{\bf T}_C = {\bf T}_3 - {\bf T}_4,  &\qquad
{\bf T}_D = {\bf T}_3 + {\bf T}_4.
\end{split}\end{align}
Note that in contrast with the generators corresponding to different lines $\{{\bf T}_1, {\bf T}_2,  {\bf T}_3,  {\bf T}_4\}$ the generators defined in eq.~(\ref{color_basis}) are not mutually commuting. The nonzero commutators amongst them are summarised by eq.~(\ref{ABCDcomm}).

Let us first consider the $n=3$ case, for which colour conservation
assumes the form
\begin{equation}
{\bf T}_3=-{\bf T}_A.
\label{colcon3}
\end{equation}
 As we saw in
eq.~(\ref{DelSp3}), the difference in eq.~(\ref{delcoll}) reduces to the single term $\Delta_3^{(3)}$ when $n=3$. In
addition, for this number of partons the second and third lines in eq.~(\ref{eq:expected_Delta})
are not present, as they consist of sums over four particle
combinations only. We thus have
\begin{align}
\Delta^{(3)}_3 ({\bf T}_3,{\bf T}_1,{\bf T}_2) &= -16 C f_{abe} f_{cde} \Big( \big\{{\bf T}_1^a, {\bf T}_1^d \big\} {\bf T}_2^b {\bf T}_3^c + \big\{{\bf T}_2^a, {\bf T}_2^d \big\} {\bf T}_1^b {\bf T}_3^c +\big\{{\bf T}_3^a, {\bf T}_3^d \big\} {\bf T}_1^b {\bf T}_2^c \Big). \label{Delta_3_legs}
\end{align}
One can put this in a form where colour conservation is made explicit using eq.~(\ref{colcon3}), after
commuting all factors of ${\bf T}_3$ to the right. Making
liberal use of the colour identities listed in
appendix~\ref{app:colid}, one can put this in the form
\begin{align}
\Delta^{(3)}_3 (-{\bf T}_1-{\bf T}_2,{\bf T}_1,{\bf T}_2)
 &= - 6 C f_{abe} f_{cde} \big\{ {\bf T}_A^a, {\bf
  T}_A^d \big\} \big\{ {\bf T}_B^b, {\bf T}_B^c \big\} + 3 C N_c^2
      {\bf T}_A \cdot {\bf T}_A, \label{gamma_split}
\end{align}
where $C$ is the part of our ansatz defined in eq.~\eqref{ansatz2}. As the right-hand side of this equation is already a constant and depends only on the
colour degrees of freedom of particles 1 and 2, it does not directly
constrain our ansatz. However, we can now require that the right hand side of eq.~\eqref{delcoll} 
equals this expression when evaluated for any number of partons $n$.

With this in mind, we now turn to the $n=4$ case. Using the definitions in eq.~(\ref{color_basis}) again, colour conservation can be written as
\begin{equation}
{\bf T}_D=-{\bf T}_A\,.
\label{colcon4}
\end{equation}
Moreover, the second term in
eq.~(\ref{delcoll}) becomes ({\it cf.} eq.~(\ref{Delta_3_legs}))
\begin{align}
\Delta^{(3)}_3 ({\bf T}_1+{\bf T}_2,{\bf T}_3,{\bf T}_4)\
&= -16 C f_{abe} f_{cde} \Big[ \big\{{\bf T}_1^a + {\bf T}_2^a, {\bf T}_1^d + {\bf T}_2^d \big\} {\bf T}_3^b {\bf T}_4^c  \\ \nonumber
&\hspace{1.4cm} + \big\{{\bf T}_3^a, {\bf T}_3^d \big\} \left( {\bf T}_1^b + {\bf T}_2^b \right) {\bf T}_4^c +\big\{{\bf T}_4^a, {\bf T}_4^d \big\} \left( {\bf T}_1^b +{\bf T}_2^b \right) {\bf T}_3^c \Big]\,.
\end{align}
After expressing all colour generators in terms of the basis of
eq.~(\ref{color_basis}), one may commute all factors of ${\bf T}_D$ to
the right, and then implement colour conservation according to eq.~(\ref{colcon4}). Using the identities in appendix~\ref{app:colid} one then obtains
\begin{equation}
\Delta^{(3)}_3 (-{\bf T}_3-{\bf T}_4,{\bf T}_3,{\bf T}_4)
 = -6 C f_{abe} f_{cde} \big\{ {\bf T}_A^a, {\bf T}_A^c \big\} \big\{ {\bf T}_C^b , {\bf T}_C^d \big \} +3 C N_c^2 {\bf T}_A \cdot {\bf T}_A\,. \label{independent_color_1}
\end{equation}
To put the first term in eq.~(\ref{delcoll}) into a form that manifests colour conservation, we first use the Jacobi identity to rewrite the expression in eq.~(\ref{eq:expected_Delta}) for general $\mathcal{F}(\rho_{1234}, \rho_{1432})$ and $C$ as
\begin{align} \label{collinear_expression}
\Delta_4^{(3)} & =- 16 C f_{abe} f_{cde} \hspace{-1cm} \sum_{\{i,j,k\} \in \{1,2,3,4\} | j<k} \hspace{-1cm} \big\{{\bf T}_i^a, {\bf T}_i^d \big\} {\bf T}_j^b {\bf T}_k^c   \\
&\hspace{0.8cm}+ 8\, {\bf T}_1^a {\bf T}_2^b {\bf T}_3^c {\bf T}_4^d \,\Big[
 f_{abe} f_{cde}  \big( \mathcal{F}(\rho_{1234}, \rho_{1432} ) - \mathcal{F}(\rho_{1243}, \rho_{1342} ) +  2 \mathcal{F}(\rho_{1324}, \rho_{1423} )  \big)
    \nonumber \\ \nonumber
&\hspace{3.3cm} +  
\big( f_{ace} f_{bde} + f_{ade} f_{bce} \big) \big( \mathcal{F}(\rho_{1234}, \rho_{1432} ) + \mathcal{F}(\rho_{1243}, \rho_{1342} )  \big)
 \Big].
\end{align}
Again employing colour conservation after moving ${\bf T}_D$ to the
right and simplifying, the sum over colour structures in the first
term of eq.~(\ref{collinear_expression}) can be rewritten as
\begin{align}\label{independent_color_eqs}
\begin{split}
f_{abe} f_{cde} \hspace{-1cm} \sum_{\{i,j,k\} \in \{1,2,3,4\} | j<k}  \hspace{-1cm} 
 \big\{{\bf T}_i^a, {\bf T}_i^d \big\} {\bf T}_j^b {\bf T}_k^c &= \frac{1}{4} f_{abe} f_{cde} \Big( \big\{ {\bf T}_A^a, {\bf T}_A^c \big\} \big\{ {\bf T}_B^b, {\bf T}_B^d \big\} + \big\{ {\bf T}_A^a, {\bf T}_A^c \big\} \big\{ {\bf T}_C^b, {\bf T}_C^d \big\}\\
& \quad + \frac{1}{2}  \big\{ {\bf T}_B^a, {\bf T}_B^c \big\} \big\{ {\bf T}_C^b, {\bf T}_C^d \big\} \Big) 
- \frac{5}{16}  N_c^2 {\bf T}_A \cdot {\bf T}_A. 
\end{split}
\end{align}
By a similar procedure, the remaining colour structures in
eq.~(\ref{collinear_expression}) evaluate to
\begin{subequations}
\label{independent_color_eqs2}
\begin{align}
&{\bf T}_1^a {\bf T}_2^b {\bf T}_3^c {\bf T}_4^d 
 \,\,f_{abe} f_{cde}  = - \frac{1}{8} f_{abe} f_{cde} \big\{ {\bf T}_A^a, {\bf T}_A^c \big\} {\bf T}_B^b {\bf T}_C^d  - i \frac{3}{16} N_c f_{abc} {\bf T}_A^a {\bf T}_B^b {\bf T}_C^c  
 - \frac{1}{16} N_c^2 {\bf T}_B \cdot {\bf T}_C\,, 
\\
&{\bf T}_1^a {\bf T}_2^b {\bf T}_3^c {\bf T}_4^d \,\big( f_{ace} f_{bde} + f_{ade} f_{bce} \big) = \frac{1}{32} f_{abe} f_{cde} 
\Big(\big\{ {\bf T}_B^a, {\bf T}_B^c \big\} \big\{ {\bf T}_C^b, {\bf T}_C^d \big\}
- \big\{ {\bf T}_A^a, {\bf T}_A^c \big\} \big\{ {\bf T}_B^b, {\bf T}_B^d \big\}\nonumber \\
&\hspace*{6cm}
- \big\{ {\bf T}_A^a, {\bf T}_A^c \big\} \big\{ {\bf T}_C^b, {\bf T}_C^d \big\} \Big) 
+ \frac{1}{64} N_c^2  {\bf T}_A \cdot  {\bf T}_A\,.
\end{align}
\end{subequations}
Substituting these expressions back into
eq.~\eqref{collinear_expression} and subtracting
eq.~(\ref{independent_color_1}), one obtains
\begin{align} \nonumber
\Big[\Delta_4^{(3)}&-\Delta_3^{(3)}\Big]_{p_1\parallel p_2} 
=-\frac14 f_{abe}f_{cde} \times 
\\
\nonumber&\Bigg\{
\Big(\{{\bf T}_A^a,{\bf T}_A^c\}\,\{{\bf T}_B^b,{\bf T}_B^d\}
-\frac12N_c^2{\bf T}_A\cdot{\bf T}_A\Big)
\bigg[{\cal F}(\rho_{1234},\rho_{1432})
+{\cal F}(\rho_{1243},\rho_{1342})+16C\bigg]_{p_1\parallel p_2}\\
\nonumber&\qquad+\{{\bf T}_C^b,{\bf T}_C^d\}\Big(
\{{\bf T}_A^a,{\bf T}_A^c\}-\{{\bf T}_B^a,{\bf T}_B^c\}
\Big)\bigg[{\cal F}(\rho_{1234},\rho_{1432})
+{\cal F}(\rho_{1243},\rho_{1342})-8C\bigg]_{p_1\parallel p_2}\Bigg\}
\\
\nonumber&-\Big(f_{abe}f_{cde}\{{\bf T}_A^a,{\bf T}_A^c\}
{\bf T}_B^b{\bf T}_C^d+\frac{3i}{2}N_c f_{abc}{\bf T}_A^a\,{\bf T}_B^b\,
{\bf T}_C^c+\frac{1}{2}N_c^2{\bf T}_B\cdot {\bf T}_C\Big)\\
&\qquad\times 
\bigg[{\cal F}(\rho_{1234},\rho_{1432})-{\cal F}(\rho_{1243},\rho_{1342})
+2{\cal F}(\rho_{1324},\rho_{1423})\bigg]_{p_1\parallel p_2}\,.
\label{collinear_expression2}
\end{align}
As discussed above, consistency of collinear factorisation means that
the right-hand side of eq.~(\ref{collinear_expression2}) must only
depend on the degrees of freedom of particles 1 and 2. This in turn
means that dependence on ${\bf T}_C$ in the collinear limit is
forbidden, which immediately implies the constraints
\begin{align}
\Big[ \mathcal{F}(\rho_{1234}, \rho_{1432} ) - \mathcal{F}(\rho_{1243}, \rho_{1342} ) +  2 \mathcal{F}(\rho_{1324}, \rho_{1423} ) \Big]_{p_1 \parallel p_2} &= 0,
\notag \\
\Big[ \mathcal{F}(\rho_{1234}, \rho_{1432} ) + \mathcal{F}(\rho_{1243}, \rho_{1342} ) \Big]_{p_1 \parallel p_2} &= 8 C.
\label{Fcalcoll}
\end{align}
Implementing these in eq.~(\ref{collinear_expression2}) we find that the latter indeed becomes equal to eq.~(\ref{gamma_split}). Thus, the quantity $\Gamma_{\bf Sp}$
is indeed found to be universal, in that it is independent of whether
one considers three or four-parton scattering. The constraints of
eq.~(\ref{Fcalcoll}) can be recast in the form
\begin{align}
F_a(z_{1234})\Big|_{p_1 \parallel p_2}&= 0 {\rm~~~~and~~~~}F_b(z_{1234})\Big|_{p_1 \parallel p_2}= 8C\,,\label{collinear_constraint_2}
\end{align}
where we used eq.~(\ref{Fdiffs}) to write the relevant combinations of ${\cal F}$ in terms of $F(z_{1234})$ and subsequently related them to the $F_i(z)$, $i\in\{1,2,3\}$, defined in eq.~(\ref{Fcombs}), getting:  
\begin{equation}\begin{split}
F_a(z)&\,\equiv F_1(z)-F_2(z) = F_1(z) - F_1\left(z/(z-1)\right)\,,\\
F_b(z)&\,\equiv -F_3(z)=F_1(z)+F_2(z)=F_1(z) + F_1\left(z/(z-1)\right)\,,
\label{FaFb123}
\end{split}\end{equation}
where $z=z_{1234}$ and where we used eq.~(\ref{Ftrans}) to write $F_2$ in terms of $F_1$. 
The interpretation of the two conditions in eq.~(\ref{collinear_constraint_2}) becomes clear upon comparing eq.~(\ref{collinear_expression}) to eq.~(\ref{Delta3form4}): the first condition corresponds to the component in $\Delta_4^{(3)}$ which is antisymmetric in both colour and kinematics variables of lines 1 and 2, and so must clearly vanish when $p_1\parallel p_2$, while the latter corresponds to the symmetric component, which instead approaches a non-zero constant in this collinear limit.

Considering the leading behaviour of the kinematic variables in
the collinear limit $P^2=2p_1\cdot p_2\rightarrow 0$, one finds
\begin{align}
\begin{split}
z  {\bar z} &= \frac{(p_1 \cdot p_2) (p_3 \cdot p_4)}{(p_1 \cdot p_3) (p_2 \cdot p_4)} \xrightarrow{p_1 \parallel p_2} \,0, \\
(1-z)(1-{\bar z}) &= \frac{(p_1 \cdot p_4) (p_3 \cdot p_2)}
{(p_1 \cdot p_3) (p_4 \cdot p_2)} \xrightarrow{p_1 \parallel p_2} \,1,
\end{split}
\end{align}
where all phases cancel, having assumed that all particles are outgoing (as far as the cross ratios are concerned, this is equivalent to working in the Euclidean region). 
These conditions together imply
\begin{align}
z\xrightarrow{p_1 \parallel p_2} 0 \qquad \text{and}\qquad 
\bar z \xrightarrow{p_1 \parallel p_2} 0\,,
 \label{collinear_zbar}
\end{align}
in which limit $F(z)$ reduces to a polynomial in
$\log(z\bar z)$ with coefficients drawn from the space of
multiple zeta values. It is easy to see that the condition on $F_a(z)$ in eq.~\eqref{collinear_constraint_2} is always satisfied, 
and thus does not provide any constraint on the coefficients $\{a_i\}$, because
\begin{equation}
F_a(z)\Big|_{p_1 \parallel p_2} = \lim_{(z,\bar{z})\to 0}\left[F_1(z) - F_1\left(z/(z-1)\right)\right]  = 0\,.
\end{equation}
%
Conversely, plugging the ansatz of eq.~(\ref{ansatz1}) into
eq.~\eqref{collinear_constraint_2}, using eq.~(\ref{FaFb123}) and
taking the leading collinear behaviour as $p_1\parallel p_2$, we find
\begin{align} \label{collim_general}
\begin{split}
F_b(z)\Big|_{p_1 \parallel p_2}&\, = \lim_{(z,\bar{z})\to 0}\Big(- F_3(z)\Big)= - \frac{a_1}{60} \log^5 (z \bar z) - \frac{a_8}{3} \zeta_2 \log^3(z \bar z) - (4 a_7+a_{10}) \zeta_3 \log^2(z \bar z)   \\
&- 2 a_{11} \zeta_2^2 \log(z \bar z)+ 8 a_9 \zeta_2 \zeta_3 + (24 a_2-12 a_3 +8 a_4 +36 a_5 - 12 a_6 + 4 a_7) \zeta_5,
\end{split}
\end{align}
which is not generically a constant, as eq.~\eqref{collinear_constraint_2} tells us it must be.
Requiring all the non-constant terms in $F_b$ to vanish gives us the constraints
\begin{equation}
a_1=0,\qquad a_{7}=-\frac{a_{10}}{4},\qquad a_8=0,\qquad a_{11}=0\,.
\label{aconditions_coll}
\end{equation}
Moreover, we can use the condition in
eq.~\eqref{collinear_constraint_2} to place constraints on the
parameters $a_{12}$ and~$a_{13}$ that enter our ansatz for $C$ in eq.~(\ref{ansatz2}). That is, imposing the constraints in
eqs.~\eqref{aconditions_regge} and \eqref{aconditions_coll} on $F_b$
and equating it to $8C$ implies
\begin{align}
a_{12}&=3a_2-\frac{3}{2}a_3+a_4+\frac{9}{2}a_5-\frac{3}{2}a_6
+\frac{1}{2}a_7,\quad\quad
a_{13}=a_9.
\label{a12a13}
\end{align}

Here we have considered particles 1 and 2 becoming collinear such that $z\rightarrow 0$, and leading to the conditions:
\begin{equation}
F_1(z)-F_2(z)\xrightarrow{z\rightarrow 0}0,\quad \quad 
F_1(z)+F_2(z)=-F_3(z)\xrightarrow{z\rightarrow 0} 8C,
\label{12lim}
\end{equation}
corresponding to the antisymmetric and symmetric parts in eq.~(\ref{Delta3form4}) under permutation of~\{1,2\}.
We may also consider the limits in which the pair of particles \{1,3\} or \{1,4\}
become collinear, implying $z\rightarrow \infty$ and
$z\rightarrow 1$ respectively.\footnote{The remaining collinear limits
  in which particles \{2,3\}, \{2,4\} or \{3,4\} become collinear correspond
  to the same limits of $z$, and thus provide no complementary
  information.} The above arguments can be repeated to show that in
the limit $p_1\parallel p_3$, one obtains the conditions
\begin{equation}
F_1(z)-F_3(z)\xrightarrow{z\rightarrow\infty} 0,\quad\quad 
F_1(z)+F_3(z)=-F_2(z)\xrightarrow{z\rightarrow\infty} 8C,
\label{13lim}
\end{equation}
and similarly if $p_1\parallel p_4$:
\begin{equation}
F_2(z)-F_3(z)\xrightarrow{z\rightarrow 1} 0,\quad\quad  
F_2(z)+F_3(z)=-F_1(z)\xrightarrow{z\rightarrow 1} 8C.
\label{14lim}
\end{equation}
The conditions of eqs.~(\ref{13lim}) and~(\ref{14lim}) can be seen to
coincide with eq.~(\ref{12lim}) upon using the transformations of
eq.~(\ref{zperms}), or equivalently they can be simply deduced from eq.~(\ref{12lim}) 
using the relations in eq.~(\ref{Ftrans}).
Thus, the additional collinear limits provide no
complementary information. In summary, the collinear limit provides
the following constraints on the parameters~$\{a_i\}$:
\begin{align}
(a_1,a_7,a_8,a_{11},a_{12},a_{13})=
\left(0,-\frac{a_{10}}{4},0,0,
3a_2-\frac{3a_3}{2}+a_4+\frac{9a_5}{2}-\frac{3a_6}{2}
+\frac{a_7}{2}, a_9\right).
\label{aconditions_coll2}
\end{align}


\section{Discussion}
\label{sec:discuss}

In the previous two sections, we have derived separate constraints on
the parameters $\{a_i\}$ entering the ans\"{a}tze of
eqs.~(\ref{ansatz1}) and~(\ref{ansatz2}), from both the Regge and
collinear limits, as summarised in eqs.~(\ref{aconditions_regge})
and~(\ref{aconditions_coll2}) respectively. We can now combine them
into a single set.  In doing so, we see that the first two conditions
in eq.~(\ref{aconditions_coll2}) are already contained in
eq.~(\ref{aconditions_regge}), but that the remaining ones are
complementary. Upon implementing the full set of constraints, the
ans\"atze of eqs.~(\ref{ansatz1}) and~(\ref{ansatz2}) reduce to
\begin{equation}
F(z)=a_4\left({\cal L}_{10101}+2\zeta_2({\cal L}_{100}+{\cal L}_{001})
\right) {\rm~~~and~~~}
C = a_4(\zeta_5+2\zeta_2\zeta_3)\,.
\label{ansatz4}
\end{equation}
We see that both $F$ and $C$ have been uniquely determined by our bootstrap procedure,
up to an overall rational number $a_4$. Since $\Delta_n^{(3)}$ depends linearly on $F$ and $C$,
the form of the three-loop correction to the dipole formula is fixed completely by symmetries and physical constraints.
Comparing the expressions in eq.~\eqref{ansatz4} with the result of ref.~\cite{Almelid:2015jia} (quoted here in
eqs.~(\ref{Fdef}) and~(\ref{Cdef})), we see that the latter can be
reproduced by setting $a_4 = 1$. Our
analysis is a highly non-trivial cross-check of the results of
ref.~\cite{Almelid:2015jia}, as well as the consistency of the bootstrap procedure. 
This is the first time that a bootstrap procedure has been successfully applied to a quantity in non-planar perturbative gauge theory. 

It is interesting to observe the complementary nature of the Regge and collinear limits used here. While the number of constraints that arise from the Regge limit is larger -- notably because it provides information on both the real and imaginary parts of the amplitude at each logarithmic order -- it is the collinear limit that relates the function $F$ and the constant $C$. This non-trivial interplay between the four- and three-line structures is a manifestation of colour conservation, or the gauge invariance of the theory. In principle such a relation could arise from the Regge limit as well, but it requires extending computations along the lines of refs.~\cite{Caron-Huot:2013fea,Caron-Huot:2017fxr} to higher logarithmic accuracy.

Given the homogeneous nature of all our collinear and Regge
constraints, we cannot use them to fix the overall
normalisation of $\Delta_n^{(3)}$. Had it not been computed, one could consider fixing this constant by various other means.
For example, it could be determined by
numerical integration at suitably high precision at a single kinematic point. Other options include analytically computing the simpler quantity $\Delta^{(3)}_3$, which is just a constant, or extracting the overall normalization factor from the three-loop four-point amplitude in ${\cal N}=4$ SYM~\cite{Henn:2016jdu}.  Alternatively, $a_4$ could be fixed by the knowledge of a non-homogeneous constraint, such as the computation of the $i\pi\,\alpha_s^3\log(s/(-t))$ term in the Regge limit of 2-to-2 scattering, which is known to receive non-vanishing contributions from $\Delta_4^{(3)}$~\cite{Caron-Huot:2017fxr}.

The fact that the form of $\Delta_n^{(3)}$ is fixed by symmetries and dynamic
constraints begs the question as to whether the same is true at higher
loop orders. Before such a program can be carried out, however, one
must take into account the fact that the structure of the soft
anomalous dimension becomes more complicated starting at four loops. This is due to the breakdown of Casimir scaling in the cusp
anomalous dimension, which was recently
demonstrated in ref.~\cite{Boels:2017skl} and requires modifying the dipole formula starting at four loops. The contribution to the soft anomalous dimension
depending only on conformally-invariant cross ratios will still be amenable to the methods employed here. In particular, we have argued
that this contribution can be expressed in terms of single-valued polylogarithms to all loop orders. However, starting at four loops $\Delta_n$ also becomes 
sensitive to the matter content of the theory, meaning that the assumptions of uniform transcendental weight and
purity will not generically apply. Additionally, as mentioned in section~\ref{sec:HPL},
single-valued polylogarithms depending on more than one
complex variable are expected to appear, since more than four Wilson lines can become correlated in higher-loop diagrams.
Even so, the types of constraints considered in this paper also generalize to higher loops. In
fact, the requirement that the amplitude factorizes in two-particle collinear limits can be directly applied at any loop order~\cite{Bern:1999ry,Kosower:1999xi,Feige:2014wja,Catani:2003vu}. 
Further computations are required to extend the types of Regge constraints we have used in the present paper to higher loops, but there exists a
well-defined framework to study the Regge limit of QCD amplitudes at arbitrary
order~\cite{Caron-Huot:2013fea,Caron-Huot:2016tzz,Caron-Huot:2017fxr}, making it possible for the relevant computations to be carried out. In particular, it is possible to explore constraints coming from multi-Regge limits~\cite{DelDuca:2011ae,DelDuca:2011xm} in addition to single-Regge limits. Lastly, one can consider multi-particle collinear limits, in which higher-particle Mandelstam invariants go on shell. The multi-particle limits of the three-loop soft anomalous dimension have already been computed, and do not give rise to any constraints on $\Delta_n^{(3)}$ beyond those implied by two-particle collinear limits~\cite{sebastian}.
However, they could provide additional information at higher
loop orders.

Finally, it would be interesting to see if a similar bootstrap procedure could be applied 
to other physical quantities of interest. The extension of our work to the massive three-loop
soft anomalous dimension is currently hampered by our lack of understanding of the corresponding space of functions, i.e.,
iterated integrals on hyperbolic 3-space, as discussed in section~\ref{sec:HPL}. Conversely, the single-emission soft current is known to be expressible in terms of SVHPLs through two loops from the calculations of ref.~\cite{Catani:2000pi,Duhr:2013msa,Li:2013lsa,ZhuTalk}, suggesting that this quantity may also be amenable to bootstrap techniques. It can also be hoped that it will eventually prove possible to bootstrap QCD amplitudes themselves; however, a much better understanding of the functions appearing in these amplitudes is required before any such approach can be attempted. 

\section*{Acknowledgments}
We are grateful for Lance Dixon, Sebastian Jaskiewicz, Jeffrey Pennington and Leonardo Vernazza for stimulating discussions. 
We would like to thank the ESI institute in Vienna and the organisers of the
program ``Challenges and Concepts for Field Theory and Applications in the Era of LHC
Run-2,'' Nordita in Stockholm and the organisers of ``Aspects of Amplitudes'' over the
summer of 2016 and the CERN TH Institute ``LHC and the Standard Model: Physics and Tools''. 
CD, AM \& CDW acknowledge the hospitality of
the Higgs Centre of the University of Edinburgh at various stages of this work.
This work is supported by the European Research Council
(ERC) under the Horizon 2020 Research and Innovation Programme through the grant 637019 ``MathAm'', by the National Science Foundation under Grant No. NSF PHY11-25915, and by
the STFC Consolidated Grants ``Particle Physics at the Higgs Centre'' and ``String Theory, Gauge Theory and Duality''.

\appendix

\section{Alternative Regge limits}
\label{app:regges}

In this appendix, we give the forms of the ansatz of
eq.~(\ref{ansatz1}) in some alternative Regge limits to that
considered in section~\ref{sec:Reggelimit}, using the analytic
continuations of eqs.~(\ref{con2}) and~(\ref{con3}). We first consider
particles 1 and 3 to be incoming. Then, in the Regge limit $s_{13}\gg
-s_{14}$, the real and imaginary parts of the functions of
eq.~(\ref{Fcombs}) become (with $L\equiv\log(s_{13}/ (-s_{14}))$)
\begin{align}\label{Reggelim3}
{\rm Re}(F_1)&\xrightarrow{s_{13} \gg -s_{14}}
-\frac{8a_1}{15} L^5 
+ \left(16 a_1 - 12 a_2 + 16 a_6 + 24 a_7 - \frac{8 a_8}{3}\right) 
  \zeta_2 L^3 + (4 a_{10}+ 16 a_7) 
 \zeta_3  L^2\nonumber \\
&+ 
   (-48a_1-4a_{11}+72a_2+48a_3-48a_4+48a_5+24a_8+24a_9) \zeta_2^2 L\nonumber \\ 
& +  (12 a_{10} + 48 a_2 - 24 a_3 + 48 a_4 + 24 a_5 - 24 a_6 + 72 a_7 - 
    8 a_9) \zeta_2 \zeta_3 \nonumber \\
&+(-24a_2 + 12a_3 - 8a_4 - 36a_5 + 12a_6 - 4a_7)\zeta_5,\nonumber \\
{\rm Re}(F_2)&\xrightarrow{s_{13} \gg -s_{14}} \frac{4a_1}{15} L^5 
+\left(-16a_1+4a_2-12a_3-12a_5-36a_6-52a_7+\frac{4a_8}{3} \right)
\zeta_2 L^3\nonumber \\
&+(-2a_{10}-8a_7)\zeta_3 L^2
+(48a_1+2a_{11}+24a_2+48a_3-48a_5+48a_6+96a_7-24a_8)\zeta_2^2 L\nonumber \\
&+(-24a_{10}-24a_2-144a_7+4a_9)\zeta_2\zeta_3\nonumber \\
&+(12 a_2-6a_3+4 a_4 +18a_5-6 a_6 +2 a_7)\zeta_5,\nonumber \\
{\rm Re}(F_3)&\xrightarrow{s_{13} \gg -s_{14}} \frac{4a_1}{15}L^5
+\left(8a_2+12a_3+12a_5+20a_6+28a_7+\frac{4a_8}{3}\right)\zeta_2 L^3
+(-2a_{10}-8a_7)\zeta_3 L^2\nonumber \\
&+(2a_{11}-96a_2-96a_3+48a_4-48a_6-96a_7-24a_9)\zeta_2^2 L\nonumber \\
&+(12a_{10}-24a_2+24a_3-48a_4-24a_5+24a_6+72a_7+4a_9)\zeta_2\zeta_3\nonumber \\
&+(12a_2-6a_3+4a_4+18a_5-6a_6+2a_7)\zeta_5, 
\end{align}
and
\begin{align}\label{Reggelim4}
\frac{1}{\pi}{\rm Im}(F_1)&\xrightarrow{s_{13} \gg -s_{14}} \frac{4a_1}{3} L^4
+(-16a_1+18a_2-24a_6-36a_7+4a_8)\zeta_2L^2+(-4a_{10}-16a_7)\zeta_3 L\nonumber \\
&+\left(\frac{48a_1}{5}+2a_{11}-18a_2-24a_3+24a_4-24a_5-24a_6-36a_7
-8a_8-12a_9\right)\zeta_2^2,\nonumber \\
\frac{1}{\pi}{\rm Im}(F_2)&\xrightarrow{s_{13} \gg -s_{14}} 
\left(-\frac{4a_1}{3}-\frac{a_2}{6}-a_3-a_5-\frac{7a_6}{3}-\frac{10a_7}{3}
\right)L^4\nonumber \\
&+(16a_1+2a_2+18a_3-4a_4+2a_5+34a_6+54a_7-4a_8+2a_9)\zeta_2 L^2\nonumber \\
&+(-4a_{10}-4a_3+8a_4+4a_5-4a_6-28a_7)\zeta_3 L\nonumber \\
&+\left(-\frac{96a_1}{5}-4a_{11}-\frac{18a_2}{5}-4a_3-\frac{64a_4}{5}
+12a_5-4a_6-\frac{32a_7}{5}+16a_8\right)\zeta_2^2
  \nonumber \\
\frac{1}{\pi}{\rm Im}(F_3)&\xrightarrow{s_{13} \gg -s_{14}} 
\left(\frac{a_2}{6}+a_3+a_5+\frac{7a_6}{3}+\frac{10a_7}{3}
\right)L^4\nonumber \\
&+(-20a_2-18a_3+4a_4-2a_5-10a_6-18a_7-2a_9)\zeta_2 L^2 \\
&+(8a_{10}+4a_3-8a_4-4a_5+4a_6+44a_7)\zeta_3 L\nonumber \\
&+\left(\frac{48a_1}{5}+2a_{11}+\frac{108a_2}{5}
+28a_3-\frac{56a_4}{5}+12a_5+28a_6+\frac{212a_7}{5}
-8a_8+12a_9\right)\zeta_2^2 \nonumber
\end{align}
respectively.

Next, we consider particles 1 and 4 to be incoming, and the Regge
limit $s_{14}\gg -s_{13}$. In this case, one finds (now with $L\equiv
\log(s_{14}/(-s_{13}))$)
\begin{align}\label{Reggelim5}
{\rm Re}(F_1)&\xrightarrow{s_{14} \gg -s_{13}}  \frac{4a_1}{15} L^5 
+\left(-16a_1+4a_2-12a_3-12a_5-36a_6-52a_7+\frac{4a_8}{3} \right)
\zeta_2 L^3\nonumber \\
&+(-2a_{10}-8a_7)\zeta_3 L^2
+(48a_1+2a_{11}-48a_2+24a_3-120a_5+120a_6+168a_7-24a_8)\zeta_2^2 L\nonumber \\
&+(-24a_{10}-24a_2-144a_7+4a_9)\zeta_2\zeta_3\nonumber \\
&+(12 a_2-6a_3+4 a_4 +18a_5-6 a_6 +2 a_7)\zeta_5,\nonumber \\
{\rm Re}(F_2)&\xrightarrow{s_{14} \gg -s_{13}} 
-\frac{8a_1}{15} L^5 
+ \left(16 a_1 - 12 a_2 + 16 a_6 + 24 a_7 - \frac{8 a_8}{3}\right) 
  \zeta_2 L^3 + (4 a_{10}+ 16 a_7) 
 \zeta_3  L^2\nonumber \\
&+ 
   (-48a_1-4a_{11}+48a_3-48a_4+48a_5-288a_6-432a_7+24a_8+24a_9) \zeta_2^2 L\nonumber \\ 
& +  (12 a_{10} + 48 a_2 - 24 a_3 + 48 a_4 + 24 a_5 - 24 a_6 + 72 a_7 - 
    8 a_9) \zeta_2 \zeta_3 \nonumber \\
&+(-24a_2 + 12a_3 - 8a_4 - 36a_5 + 12a_6 - 4a_7)\zeta_5,\nonumber \\
{\rm Re}(F_3)&\xrightarrow{s_{14} \gg -s_{13}}  \frac{4a_1}{15}L^5
+\left(8a_2+12a_3+12a_5+20a_6+28a_7+\frac{4a_8}{3}\right)\zeta_2 L^3
+(-2a_{10}-8a_7)\zeta_3 L^2\nonumber \\
&+(2a_{11}+48a_2-24a_3+48a_4+72a_5+168a_6+264a_7-24a_9)\zeta_2^2 L\nonumber \\
&+(12a_{10}-24a_2+24a_3-48a_4-24a_5+24a_6+72a_7+4a_9)\zeta_2\zeta_3\nonumber \\
&+(12a_2-6a_3+4a_4+18a_5-6a_6+2a_7)\zeta_5, 
\end{align}
and
\begin{align}\label{Reggelim6}
\frac{1}{\pi}{\rm Im}(F_2)&\xrightarrow{s_{14} \gg -s_{13}} 
\left(-\frac{4a_1}{3}-\frac{a_2}{6}-a_3-a_5-\frac{7a_6}{3}-\frac{10a_7}{3}
\right)L^4\nonumber \\
&+(16a_1-16a_2+6a_3-4a_4-10a_5+22a_6+30a_7-4a_8+2a_9)\zeta_2 L^2\nonumber \\
&+(-4a_{10}-4a_3+8a_4+4a_5-4a_6-28a_7)\zeta_3 L\nonumber \\
&+\left(-\frac{96a_1}{5}-4a_{11}-\frac{108a_2}{5}+92a_3-\frac{184a_4}{5}
+12a_5-196a_6-\frac{1532a_7}{5}+16a_8 +36a_9\right)\zeta_2^2
  \nonumber \\
\frac{1}{\pi}{\rm Im}(F_2)&\xrightarrow{s_{14} \gg -s_{13}} \frac{4a_1}{3} L^4
+(-16a_1+18a_2-24a_6-36a_7+4a_8)\zeta_2L^2+(-4a_{10}-16a_7)\zeta_3 L\nonumber \\
&+\left(\frac{48a_1}{5}+2a_{11}+18a_2-24a_3+24a_4-24a_5+120a_6+180a_7
-8a_8-12a_9\right)\zeta_2^2,\nonumber \\
\frac{1}{\pi}{\rm Im}(F_3)&\xrightarrow{s_{14} \gg -s_{13}} 
\left(\frac{a_2}{6}+a_3+a_5+\frac{7a_6}{3}+\frac{10a_7}{3}
\right)L^4\nonumber \\
&+(-2a_2-6a_3+4a_4+10a_5+2a_6+6a_7-2a_9)\zeta_2 L^2 \\
&+(8a_{10}+4a_3-8a_4-4a_5+4a_6+44a_7)\zeta_3 L\nonumber \\
&+\left(\frac{48a_1}{5}+2a_{11}+\frac{18a_2}{5}
-68a_3-\frac{64a_4}{5}+12a_5+76a_6+\frac{632a_7}{5}
-8a_8-24a_9\right)\zeta_2^2 \nonumber.
\end{align}

\section{Useful colour identities}
\label{app:colid}

Here, we collect a number of colour algebra identities that are useful
when applying the collinear constraint in
section~\ref{sec:collconstrain}. Contraction of a pair of ${\rm SU}(N_c)$ structure
constants gives
\begin{equation}
f_{acd}f_{bcd}=N_c\delta_{ab}\,.
\label{colour1}
\end{equation}
Then the Jacobi identity of eq.~(\ref{Jacobi}) implies
\begin{equation}
f_{ade}f_{beg}f_{cgd}=\frac{N_c}{2}f_{abc}.
\label{colour2}
\end{equation}
We also make use of a number of useful identities resulting from
antisymmetry of the structure constants under interchange of any two
indices:
\begin{align}\label{colour3}
\begin{split}
f_{cde} \{ {\bf T}_i^a, {\bf T}_i^d \} A^{bc} &= 0 \\
f_{abc} {\bf T}_i^a {\bf T}_i^b &= \frac 1 2 f_{abc}[ {\bf T}_i^a, {\bf T}_i^b] \\
f_{abe} f_{cde} f_{h(x) h(y) g} \{ {\bf T}_i^{h(z)}, {\bf T}_i^g \} &= 0 ,
\end{split}
\end{align}
where $A^{bc}=-A^{cb}$ is an arbitrary antisymmetric matrix of
dimension $(N_c^2-1)$, and the last identity holds for any invertible
map $h: \{ x, y, z \} \mapsto \{a, b, c, d\}$. 

Finally, the non-zero commutators among the colour operators defined in eq.~\eqref{color_basis} are
\begin{align}
[{\bf T}_A^a, {\bf T}_A^b ] &= i f^{abc} {\bf T}_A^c, \quad &[{\bf T}_B^a, {\bf T}_B^b ] &= i f^{abc} {\bf T}_A^c, \nonumber \\
[{\bf T}_A^a, {\bf T}_B^b ] &= i f^{abc} {\bf T}_B^c, \quad &[{\bf T}_B^a, {\bf T}_A^b ] &= i f^{abc} {\bf T}_B^c, \nonumber \\ 
[{\bf T}_C^a, {\bf T}_C^b ] &= i f^{abc} {\bf T}_D^c, \quad &[{\bf T}_D^a, {\bf T}_D^b ] &= i f^{abc} {\bf T}_D^c, \nonumber \\
[{\bf T}_C^a, {\bf T}_D^b ] &= i f^{abc} {\bf T}_C^c, \quad &[{\bf T}_D^a, {\bf T}_C^b ] &= i f^{abc} {\bf T}_C^c.
\label{ABCDcomm}
\end{align}
These are useful for applying colour conservation in this basis.

\bibliography{refs.bib}

\end{document}